\documentclass[aps,prd,10pt,twocolumn,preprintnumbers,superscriptaddress,letterpaper,longbibliography]{revtex4-2}

\usepackage{graphicx}
\usepackage[caption=false]{subfig} 
\usepackage[export]{adjustbox} 
\usepackage{amsmath, amssymb}
\usepackage{hyperref}
\hypersetup{
  colorlinks  = true,
  urlcolor    = blue,
  linkcolor   = blue,
  citecolor   = red
}
\usepackage[capitalize]{cleveref}
\usepackage{makecell}
\usepackage{multirow}
\newcommand{\orcid}[1]{\begingroup
  \hypersetup{hidelinks}\href{https://orcid.org/#1}{\includegraphics[width=10pt]{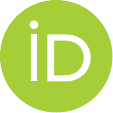}} \endgroup}
\usepackage{booktabs}

\begin{document}

\title{Milky Way satellite velocities reveal the Dark Matter power spectrum at small scales}

\author{Ivan Esteban \orcid{0000-0001-5265-2404}\,}
\email{esteban.6@osu.edu}
\affiliation{Center for Cosmology and AstroParticle Physics
  (CCAPP), Ohio State University, Columbus, Ohio 43210}
\affiliation{Department of Physics, Ohio State University, Columbus, Ohio 43210}

\author{Annika H.~G.~Peter \orcid{0000-0002-8040-6785}\,}
\email{peter.33@osu.edu}
\affiliation{Center for Cosmology and AstroParticle Physics
  (CCAPP), Ohio State University, Columbus, Ohio 43210}
\affiliation{Department of Physics, Ohio State University, Columbus, Ohio 43210}
\affiliation{Department of Astronomy, Ohio State University, Columbus, Ohio 43210}
\affiliation{School of Natural Sciences, Institute for Advanced Study, 1 Einstein Drive, Princeton, New Jersey 08540}

\author{Stacy Y.~Kim \orcid{0000-0001-7052-6647}\,}
\email{s.kim@surrey.ac.uk}
\affiliation{Department of Physics, University of Surrey, Guildford, GU2
7XH, United Kingdom}

\date{\today}


\begin{abstract}
Dark Matter (DM) properties at small scales remain uncertain. Recent theoretical and observational advances have provided the tools to narrow them down. Here, we show for the first time that the correlation between internal velocities and sizes of dwarf galaxies is a sharp probe of small-scale DM properties. We study modified DM power spectra, motivated by DM production during inflation. Using semi-analytic models and scaling relations, we show that such models can change the kinematics and structure of dwarf galaxies without strongly affecting their total abundance. We analyze data from Milky Way classical satellite galaxies and those discovered with the Sloan Digital Sky Survey (SDSS), finding that the DM power spectrum at comoving scales ${4\, \mathrm{Mpc}^{-1} < k < 37\,\mathrm{Mpc}^{-1}}$ cannot deviate by more than a factor of $\sim 2.5$ from scale invariance. Our results are robust against baryonic uncertainties such as the stellar mass-halo mass relation, halo occupation fraction, and subhalo tidal disruption; allowing us to independently constrain them. This work thus opens a window to probe both dwarf galaxy formation models and small-scale DM properties.
\end{abstract}

\maketitle


\section{Introduction}
\label{sec:intro}
Dark Matter (DM) is the second most abundant component of the Universe, yet its nature remains mysterious. At scales well above the size of a galaxy, observations are compatible with a cold, collisionless particle~\cite{Planck:2018vyg, DES:2021wwk, SDSS:2017yll}: the Cold Dark Matter (CDM) paradigm. At smaller scales, this paradigm has been challenged in the past by discrepancies between simulations and observations~\cite{Bullock:2017xww, Buckley:2017ijx,crnojevic2021,sales2022}.

These ``small-scale'' tensions have sparked interest in DM models that may alleviate them by affecting structure formation at galactic and sub-galactic scales~\cite{Tulin:2017ara, Dodelson:1993je, Ellis:1983ew, Hu:2000ke}. Even if baryonic physics and improved observations may also alleviate them~\cite{Bullock:2017xww, Buckley:2017ijx,brooks2019,sales2022}, the interest of modified DM models is broader, as they provide a framework to characterize the properties of DM. Theoretical and observational techniques developed to understand these tensions are then an effective tool to establish the nature of DM. Previous work used Milky Way satellite luminosities~\cite{Tollerud:2008ze, Horiuchi:2013noa, kennedy2014, Kim:2017iwr, Jethwa:2016gra, DES:2020fxi, dekker2022, DES:2022doi, Akita:2023yga} and stellar velocity dispersions~\cite{Kim:2021zzw, Akita:2023yga} to constrain the DM particle mass and interactions.

In this paper, we develop a powerful observable to determine small-scale DM properties: a joint constraint of the internal velocity, size, and total abundance of dwarf galaxies. Using semi-analytic techniques and scaling relations, inspired by physics models and calibrated against simulations, we show that the correlation between these properties contains information on DM physics that can be disentangled from galaxy formation uncertainties. We perform a statistical analysis of classical and SDSS Milky Way dwarfs, and quantify the agreement with CDM.

\begin{figure}[b]
    \centering
    \vspace*{-0.2cm}
    \includegraphics[width=\columnwidth]{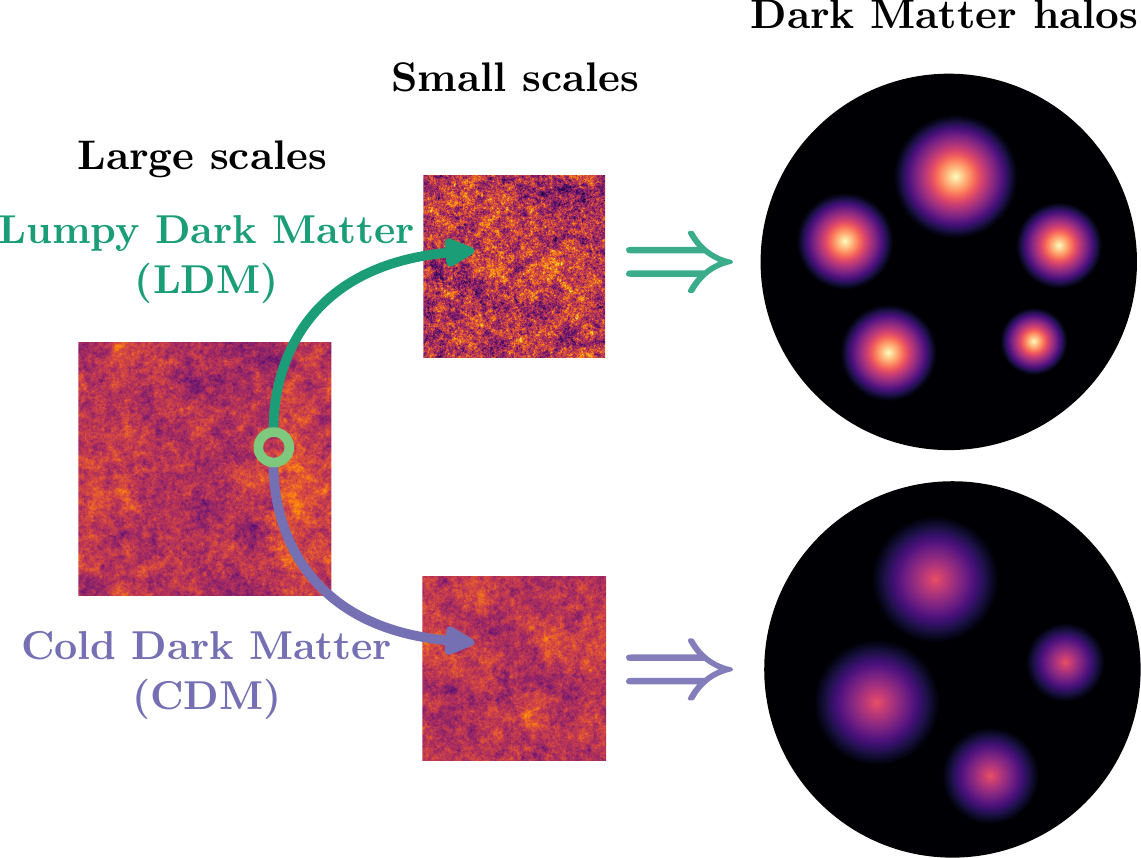}
    
    \caption{Depiction of Lumpy Dark Matter models, generated with \href{https://github.com/garrett361/cmbpy}{cmbpy}. \emph{Due to the enhanced power at small scales, in such models halos are more dense and abundant than in CDM}.}
    \vspace*{-0.7cm}
    \label{fig:cartoon}
\end{figure}

The method broadly applies to DM models with modified structure formation or primordial power spectra~\cite{Tulin:2017ara, Dodelson:1993je, Ellis:1983ew, Hu:2000ke, Subramanian:2015lua, Achucarro:2022qrl}. In this first paper, we mainly focus on models with enhanced structure at small scales. This is a natural prediction if DM is produced out of quantum fluctuations during inflation~\cite{Graham:2015rva, Alonso-Alvarez:2018tus, Tenkanen:2019aij}. Even if these models do not require non-gravitational interactions between DM and baryons, limiting the detectability of DM in the laboratory, DM being produced out of small-scale isocurvature fluctuations entails significant substructure that may be tested. Similar effects are also predicted by primordial magnetic fields~\cite{Subramanian:2015lua} or in non-standard inflationary models~\cite{Zentner:2002xt, Zentner:2003yd, Achucarro:2022qrl}. Because of the enhanced substructure, we name these DM models \emph{Lumpy Dark Matter (LDM)}. 

\Cref{fig:cartoon} previews the big picture. LDM models resemble CDM on large scales, but on small scales density fluctuations are greater. Hence, when these fluctuations collapse to form DM halos, the larger fluctuation amplitudes in LDM result in more halos than in CDM. LDM halos also form earlier, when the Universe is denser, so their internal densities are higher. Our analysis probes both the increased abundance and densities, allowing us to \emph{determine the DM power spectrum with good precision at small scales}, where to our knowledge only strong lensing determinations exist~\cite{Gilman:2021gkj}. As usual in the literature, in this first study we use semi-analytic techniques~\cite{Kamionkowski:1999vp, Zentner:2002xt, Sabti:2021unj, Gilman:2021gkj}. Future work could improve on these results using simulation-based inference, ideally including baryons.

Quantitatively, LDM models are characterized by an enhanced primordial curvature power spectrum at small scales, that we parametrize as
\begin{equation}
    \mathcal{P}_\mathcal{R}(k) \propto k^{n_s-1} \left[1 + \left(k / k_\mathrm{cut}\right)^{n_\mathrm{cut} - n_s}\right] \, ,
    \label{eq:powerSpec}
\end{equation}
with $k$ comoving wavenumber, $n_s \simeq 0.97$ the standard CDM spectral index~\cite{Planck:2018vyg}, $k_\mathrm{cut}$ the comoving wavenumber above which power is enhanced, and $n_\mathrm{cut} > n_s$ the enhanced spectral index. This expression interpolates from $\mathcal{P}_\mathcal{R} \propto k^{n_s-1}$ at $k \ll k_\mathrm{cut}$ to $\mathcal{P}_\mathcal{R} \propto k^{n_\mathrm{cut}-1}$ at $k \gg k_\mathrm{cut}$. Perturbations evolve as in CDM~\cite{Graham:2015rva, Alonso-Alvarez:2018tus, Tenkanen:2019aij}.

LDM models are phenomenologically interesting because, in contrast to most of the modified DM models studied in the literature~\cite{Tulin:2017ara, Dodelson:1993je, Ellis:1983ew, Hu:2000ke}, the DM power spectrum is enhanced instead of suppressed. This opens new observational effects and channels with which to test them. In addition, some datasets suggest a ``too many satellites'' problem~\cite{Dalal:2001fq, Kim:2017iwr, 2019MNRAS.488.4585G} that might point to such models.

The rest of the paper is organized as follows. \Cref{sec:consequences_DM} quantifies the impact of LDM models on DM halo properties. \Cref{sec:consequences_MW} links to the properties of dwarf galaxies, discusses baryonic effects, and shows that the correlation between galaxy kinematics and size is a sharp probe of DM substructure. \Cref{sec:analysis} describes our statistical analysis and results, and comments on suppressed power. \Cref{sec:conclusions} concludes and points out future research directions. The Appendices provide further details.

Below, we assume a Planck 2018 cosmology~\cite{Planck:2018vyg} ($h = 0.67$, $\Omega_\mathrm{m} = 0.31$, $\sigma_8 = 0.81$, $\Omega_\Lambda = 0.69$, $\Omega_\mathrm{b} = 0.05$). We carry out DM calculations with the \texttt{Galacticus}~\cite{Benson:2010kx} semi-analytic halo formation code. Unless otherwise specified, we evaluate Milky Way satellite properties at median infall redshift $z_\mathrm{infall} = 1$~\cite{Rocha:2011aa, Dooley:2016xkj, 2019arXiv190604180F}, i.e., when entering the Milky Way virial radius. We define halo virial masses as $M_{200c}$, i.e., the mean overdensity within the virial radius relative to the critical density is $\Delta = 200$. We assume a Milky Way mass of $10^{12} \, M_\odot$ and concentration of $9$~\cite{Cautun:2019eaf}. We comment on these choices in \cref{sec:consequences_MW}.

\section{Impact on dwarf galaxy halos}
\label{sec:consequences_DM}

In this section, we use semi-analytic techniques to compute the impact of LDM models on DM halos. We show that enhanced power at small scales increases the number of low-mass halos and their central densities, leading to unique observational signatures in dwarf galaxies.

\subsection{Formalism}
\label{sec:consequences_DM_formalism}
We follow the extended Press-Schechter formalism~\cite{Press:1973iz, Bond:1990iw}, which assigns DM halos to regions where density perturbations exceed a threshold. This formalism is expected to hold for a wide range of power spectra and cosmological models~\cite{Tinker:2008ff, Benson:2012su}, and connects DM halos to the power spectrum through the variance of the density field
\begin{equation}
    \sigma^2(R(M), z) \equiv \int \frac{\mathrm{d}^3 k}{(2 \pi)^3} P_\mathrm{lin}(k, z) |W(k, R)|^2 \, ,
    \label{eq:sigmasq}
\end{equation}
where $R(M) \equiv \left[ 3 M/(4\pi \bar{\rho}_m)\right]^{1/3}$ is the Lagrangian radius of the halo with $M$ its mass and $\bar{\rho}_m$ the average matter density of the Universe, $z$ is redshift, $P_\mathrm{lin}$ is the linear matter power spectrum, and $W(k, R)$ is a window function that connects comoving wavenumber to Lagrangian radius. To avoid spurious halos at scales larger than those with enhanced power, we use a sharp $k$-space window~\cite{Benson:2012su}
\begin{equation}
    W(k, R) = \theta\left(2.5/R - k\right) \, ,
    \label{eq:window}
\end{equation}
with $\theta$ the step function. Different window functions do not significantly change the results~\cite{Leo:2018odn}. 

The number density of halos per unit mass, i.e., the halo mass function; is then
\begin{equation}
\frac{\mathrm{d}n_\mathrm{halo}}{\mathrm{d}M} = f(\sigma, z) \frac{\bar{\rho}_m}{M} \left| \frac{\mathrm{d}\log \sigma(R(M), z)}{\mathrm{d}M}\right| \, ,
\label{eq:hmf}
\end{equation}
we use the Sheth-Tormen mass function $f$, that extends the Press-Schechter formalism to ellipsoidal collapse~\cite{Sheth:2001dp}
\begin{equation}
    f(\sigma, z) = A \left[1 + (a \nu)^{-p} \right] \sqrt{a} \, e^{-a \nu/2}/\sqrt{2\pi\nu} \, ,
    \label{eq:SThmf}
\end{equation}
where $\nu \equiv \delta_c/\sigma$ with $\delta_c$ the linear overdensity threshold for spherical collapse; and $A$, $a$, and $p$ are parameters calibrated to simulations. This formalism reproduces simulations with modified power spectra~\cite{Schneider:2013ria, Hirano:2015wla, Hirano:2023auh}, and CDM simulations down to halo masses $M \sim 10^7 M_\odot$~\cite{Bohr:2021bdm}. Lower-mass halos are not expected to host galaxies and hence are not relevant for our analysis (see \cref{sec:satellites_formalism}). We set $\delta_c = 1.686$, corresponding to an Einstein-de Sitter Universe; and $A=0.30$, $a=0.79$, and $p=0.22$~\cite{2019MNRAS.485.5010B}.  As we are interested in Milky Way satellites, we need the \emph{subhalo} mass function, which we compute following the merger tree algorithm from Refs.~\cite{Cole:2000ex, Parkinson:2007yh, Sheth:1999su}. This algorithm describes the hierarchical formation history of DM halos. As we show below, our results do not strongly rely on the halo mass function, so potential uncertainties related to the extended Press-Schechter formalism in LDM models are subdominant contributions to our error budget. We include an overall 20\% suppression of the subhalo mass function due to baryons across all relevant masses ($M < 10^{11} M_\odot$, see below)~\cite{2017MNRAS.471.1709G, Benson:2019jio, 2020MNRAS.492.5780R, Samuel:2019ylk}.

We also explore halo density profiles. For low-mass halos, we assume an NFW form~\cite{Navarro:1996gj}
\begin{equation}
\label{eq:NFW}
    \rho(r) = \frac{\rho_s}{r/r_s \left(1 + r/r_s\right)^2} \, ,
\end{equation}
with $\rho_s$ a characteristic density and $r_s$ the scale radius. For higher-mass halos, we include baryonic feedback by making the profiles cored as we describe in \cref{sec:satellites_formalism}. We parametrize $r_s$ in terms of the concentration
\begin{equation}
    c_{200c} \equiv r_\mathrm{vir} / r_s \, ,
\end{equation}
with $r_\mathrm{vir}$ the virial radius. For fixed mass, halos with larger concentrations have larger central densities. 

As already discussed in the original NFW paper, there is a correlation between halo concentration and formation time, that generates a mass-concentration relation. \emph{Halos formed at higher redshift have higher concentrations at the present, reflecting that the average matter density of the Universe was larger at formation}~\cite{Navarro:1996gj, Wechsler:2001cs, Ludlow:2016ifl, Diemer:2018vmz}. This correlation was studied in detail by Diemer and Joyce~\cite{Diemer:2018vmz}, who proposed a semi-analytic relation between concentration, mass, power-spectrum slope, and formation redshift; not tuned to any specific cosmology or redshift range. This was shown to accurately reproduce the results of N-body simulations for a variety of masses, redshifts, and power spectra (see also Ref.~\cite{Wang:2019ftp}). Hence, we adopt the mass-concentration relation from Ref.~\cite{Diemer:2018vmz}. When computing concentrations, we include a lognormal scatter of 0.16\,dex~\cite{Diemer:2014gba}; i.e., we assume that halo concentration follows a lognormal distribution with a median given by the relation in Ref.~\cite{Diemer:2018vmz} and a standard deviation of 0.16\,dex.

\subsection{Consequences}
\label{sec:consequences_DM_consequences}

The impact of enhanced power spectra on halo properties is now more explicit. From \cref{eq:sigmasq}, enhanced power increases the variance of the density field $\sigma$, which leads to \emph{enhanced halo abundances}, see \cref{eq:hmf,eq:SThmf}. Halos also form earlier, as can be understood from the collapsed mass fraction in extended Press-Schechter theory~\cite{Lacey:1993iv, Ludlow:2016ifl},
\begin{equation}
    M_\mathrm{coll}(z)  = M_0 \operatorname{erfc}\left( \frac{\delta_c \left/ \left[\sqrt{2}(1 - D(z_0)/D(z))\right]\right.}{\sqrt{\sigma^2(f M_0, z) - \sigma^2(M_0, z)}} \right) \, ,
    \label{eq:M_coll}
\end{equation}
with $M_\mathrm{coll}$ the mass of halo progenitors more massive than $f M_0$ that have collapsed by redshift $z$, $M_0 \equiv M(z_0)$ the final halo mass, and $D(z)$ the linear growth factor. As $M_\mathrm{coll}(z)$ is a growing function of $\sigma$, enhanced power leads to early halo formation and \emph{high concentrations}~\cite{Hirano:2015wla, Zentner:2002xt}. 

Below, we quantify both effects and show that the impact on halo concentrations is more dramatic. This can greatly affect the kinematics of galaxies inhabiting them.

\begin{figure}[t]
    \centering
    \includegraphics[width=\columnwidth]{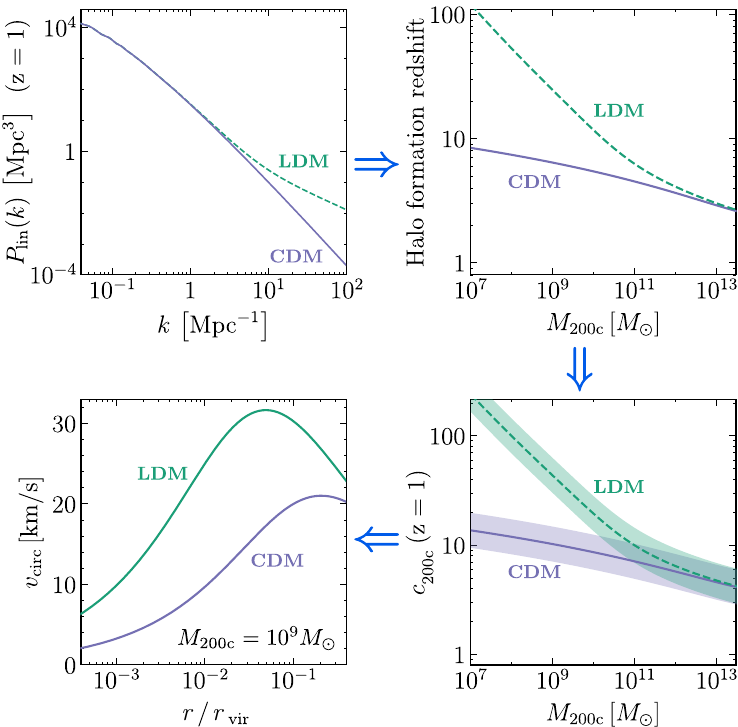}
    
    \caption{Impact of LDM models on halo structure. An enhanced power spectrum (top left) leads to \emph{early halo formation} (top right), when the DM density is higher, leading to \emph{more concentrated halos} (bottom right,  shaded regions represent $1\sigma$ scatter) and \emph{higher internal velocities} (bottom left). }
    \label{fig:physics}
\end{figure}

\Cref{fig:physics} shows that LDM models can significantly affect the internal halo kinematics due to the enhanced concentrations. For LDM, we set $k_\mathrm{cut} = 8 \, \mathrm{Mpc}^{-1}$ and $n_\mathrm{cut} = 2.6$ (our analysis below excludes this). The top-right panel shows the redshift of halo formation computed following Ref.~\cite{Ludlow:2016ifl}, defined as the redshift at which the mass within $r_s$ was contained in halo progenitors more massive than 2\% of the final halo mass. The bottom-right panel shows halo concentrations computed following Ref.~\cite{Diemer:2018vmz} as described above, along with 0.16\,dex scatter~\cite{Diemer:2014gba}. The bottom-left panel shows the circular velocity curve, $v_\mathrm{circ}(r) = \sqrt{G M(<r)/r}$ with $G$ Newton's constant, $M(<r)$ the enclosed mass, and $r$ radius; for fixed halo mass and an NFW profile. Other halos are similarly affected, but in a mass- and model-dependent way. These plots do not include tidal stripping or baryonic effects, that we introduce in \cref{sec:satellites_formalism}.

The results in \cref{fig:physics} quantify the effects described above. If power is enhanced below a given scale, halos below a given mass will form earlier. This leads to large concentrations, i.e., large DM densities close to the halo center, implying large circular velocities. 

A fortuitous cancellation increases the impact of modified power on the mass-concentration relation. For the currently favored nearly-scale-invariant power spectrum, at small scales $P_\mathrm{lin}(k) \propto k^{-3}$ approximately~\cite{Dodelson:2003ft}. The integrand in \cref{eq:sigmasq} is then proportional to $k^{-1}$, and $\sigma(R)$ only grows logarithmically with $1/R$: at small scales $\sigma(R)$ ``flattens out'', all halos form at similar redshifts, and thus have similar concentrations. Deviations from scale invariance such as those we consider induce a stronger dependence of $\sigma$ on $R$, greatly increasing the differences in formation time and concentration between halos.

\begin{figure}[t]
    \centering
    \includegraphics[width=\columnwidth]{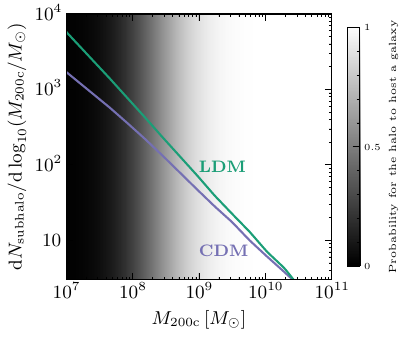}
    
    \caption{Impact of LDM models on the subhalo mass function. The grayscale shows the probability for halos to host galaxies in an illustrative model (we later set this free, see the text). \emph{The models we explore marginally affect visible subhalo abundance; our main observable is the enhanced concentrations}.}
    \label{fig:hmf}
\end{figure}

\Cref{fig:hmf} shows that visible subhalo abundance is not much enhanced in the LDM models we explore. We show the $z=0$ subhalo mass function for a parent halo with Milky-Way mass, computed with \texttt{Galacticus} following Refs.~\cite{Cole:2000ex, Parkinson:2007yh, Sheth:1999su} as described above. We use the same LDM parameters as in \cref{fig:physics}, $k_\mathrm{cut} = 8 \, \mathrm{Mpc}^{-1}$ and $n_\mathrm{cut} = 2.6$. The background shows the halo occupation fraction, i.e., the probability for a DM halo to host a galaxy, from Ref.~\cite{Dooley:2016xkj} (when performing our data analysis below, we set this free). $M_{200c}$ is the subhalo mass at infall; we discuss tidal effects and subhalo disruption in \cref{sec:satellites_formalism}.

The largest impact of LDM on halo abundances is on halos that are not expected to host galaxies: the subhalo mass function differs from CDM by $\sim 35\%$ at $10^9 \, M_\odot$ and by $\sim 50\%$ at $10^8 \, M_\odot$. Thus, our main observable is \emph{not} total halo abundance, which we include for completeness and to lift degeneracies, but internal halo properties.

Overall, we conclude that LDM models strongly affect central densities of low-mass halos. Any kinematic probe of the inner part of such halos is thus a sensitive probe of LDM models. In the following, we follow the procedure in Ref.~\cite{Kim:2021zzw}, extending the code \href{https://github.com/stacykim/disSat}{\texttt{dis}} to predict and observationally probe Milky Way satellite circular velocities and sizes, linking them to the DM power spectrum.

\section{Impact on Milky Way satellites}
\label{sec:consequences_MW}

In this section, we link low-mass DM halo properties to dwarf galaxy properties. We show that enhanced power predicts a characteristic relation between stellar kinematics and galaxy size, that can be observationally explored.

As mentioned in \cref{sec:intro}, we fix the Milky Way mass to $10^{12} \, M_\odot$ and evaluate satellite properties at median infall redshift. As we show in \cref{sec:app_statistics,sec:app_baryon}, the impact of leaving these free in our analysis is negligible, as other free parameters dominate our error budget. We also neglect specific aspects of the Milky Way merger history, like the Magellanic Clouds. Although recent work shows that they impact satellite counts~\cite{Deason:2015hla, Barry:2023ksd, 2021MNRAS.504.5270D,DES:2019ltu,2021ApJ...920L..11N}, the effect is smaller than our overall error budget (see \cref{sec:app_statistics}). In addition, we use satellites discovered by SDSS, whose footprint does not contain the Magellanic Clouds (see Ref.~\cite{2022MNRAS.516.3944M}). If other constraints improve in future work, or if satellites recently discovered in the Southern Hemisphere are included~\cite{2018ApJ...852...68C, 2018ApJ...857...70C, 2018arXiv180902259J, 2018ApJ...863...25M, DES:2020szo, DELVE:2019xvr, DELVE:2021rcc, DELVE:2021lds, DELVE:2022ijm, DELVE:2022jls, Torrealba:2018svf, 2016MNRAS.461.2212J, 2021ApJ...923..140G}, this may need revisiting.

\subsection{Formalism}
\label{sec:satellites_formalism}
It is challenging to directly observe low-mass DM halos. Avenues include gravitational lensing~\cite{Dalal:2001fq}, already quite mature (see e.g., Refs.~\cite{Vegetti:2014lqa, Inoue:2014jka, Hezaveh:2016ltk, Birrer:2017rpp, Gilman:2019nap, Hsueh:2019ynk}); perturbations to the Milky Way stellar disk~\cite{1993ApJ...403...74Q, Feldmann:2013hqa}; or disruption of stellar streams~\cite{Yoon:2010iy, Carlberg:2011xj, Aganze:2023nkp}. These may probe halos smaller than those that host galaxies. Yet observing the dwarf galaxies that low-mass halos host provides unique access to their central regions. This, as shown above in \cref{fig:physics}, is a sharp probe of LDM models. Dwarf galaxies are also DM-dominated~\cite{Tollerud:2010bj, Strigari:2008ib}, alleviating baryonic uncertainties that are more significant for more massive galaxies.

Below, we describe how we use the output from \texttt{Galacticus} (subhalo masses, concentrations, and total abundance) to build a Milky Way satellite galaxy population. To probe the central kinematics affected by LDM models, we need to predict the relationship among galaxy sizes, internal velocities, and observational abundance.

We quantify galaxy size by its 2D-projected half-light radius~\cite{Kim:2021zzw}, that we denote as $R_\mathrm{eff}$. It is correlated with its stellar mass $M_*$ by an empirical scaling relation,
\begin{equation}
    \log_{10} \left(R_\mathrm{eff}/\mathrm{kpc}\right) = 0.268 \log_{10} \left(M_*/M_\odot\right) - 2.11 \, . 
    \label{eq:R_eff}
\end{equation}
This relation has 0.234\,dex lognormal scatter. It was obtained in Ref.~\cite{Kim:2021zzw} using isolated dwarf data~\cite{McConnachie:2012vd, Read:2017lvq}, and it is similar to the relation found for satellites in the Local Group~\cite{Danieli:2017uvz} and in simulations~\cite{Jiang:2018ioo}. Although this relation can have uncertainties at low $R_\mathrm{eff}$ due to selection effects related to surface brightness, these effects impact faint galaxies with a very low observation probability (see below), which are a subdominant sample in our analysis (see also \cref{sec:app_statistics}).

The stellar mass $M_*$, in turn, links to the infall halo mass~\cite{Moster:2012fv} $M_{200c}$ via the stellar mass-halo mass relation, that we parametrize as a power law
\begin{equation}
    M_* = M_{200c} \, \mathcal{N} \left(\frac{M_{200c}}{M_0}\right)^{\beta^{M_*}}  \, ,
    \label{eq:Mstar}
\end{equation}
with $\mathcal{N}$ a constant, $M_0$ a reference mass, and $\beta^{M_*}$ the power-law slope. At dwarf galaxy masses this relation is poorly constrained, and most relations are calibrated at higher mass and then extrapolated~\cite{Moster:2012fv, Dooley:2016xkj, Wechsler:2018pic, 2021ApJ...923...35M} (see, however, Refs.~\cite{2023MNRAS.519..871Z, 2022MNRAS.516.3944M, DES:2019ltu}). At $z=1$,   Ref.~\cite{Moster:2012fv} found $\mathcal{N} = 0.046$, $M_0 = 1.5\times 10^{12} M_\odot$, and $\beta^{M_*} = 0.96$. The scatter in this relation is also uncertain and potentially mass-dependent~\cite{Garrison-Kimmel:2016szj, 2019MNRAS.483.1314B, Grand:2021fpx, 2021ApJ...923...35M}; we parametrize it as
\begin{equation}
    \sigma(M_*) = \sigma^{M_*} + \gamma^{M_*} \log_{10} \frac{M_{200 c}}{10^{11}\, M_\odot} \,\mathrm{dex}\, .
    \label{eq:Mstar_scatter}
\end{equation}
In our data analysis below, we set $\beta^{M_*}$, $\sigma^{M_*}$, and $\gamma^{M_*}$ to be free parameters. This also allows us to observationally explore the low-mass stellar mass-halo mass relation.

We quantify internal velocity by the line-of-sight stellar velocity dispersion $\sigma_\mathrm{los}^*$. We use the estimator from~\cite{Wolf:2009tu}
\begin{equation}
    \sigma_\mathrm{los}^* = \sqrt{\frac{G}{4} \frac{M(<r_{1/2})}{R_\mathrm{eff}}} \, ,
    \label{eq:sigma_los}
\end{equation}
with $M(<r_{1/2})$ the mass enclosed by the 3D half-light radius $r_{1/2} =  R_\mathrm{eff}/0.75$ (this is accurate for a wide variety of stellar distributions~\cite{Wolf:2009tu}). Other estimators produce very similar results~\cite{Kim:2021zzw}, see \cref{sec:app_baryon}. $M(<r)$ only includes DM, as the stellar mass is subdominant~\cite{Tollerud:2010bj, Strigari:2008ib}.

$\sigma_\mathrm{los}^*$ links to halo concentration through the density profile entering $M(<r)$, which can be affected by baryonic feedback that turns central NFW ``cusps'' predicted by DM-only simulations into ``cores''~\cite{2005MNRAS.356..107R, Leaman:2012bi, Weisz:2011mw, 2012MNRAS.422.1231G, Teyssier:2012ie, 2014MNRAS.437..415D, Kauffmann:2014cda, McQuinn:2015pba, Penarrubia:2012bb, Errani:2022aru} (this is expected to happen at halo masses $M\lesssim 10^8$--$10^{10}$, although there are large uncertainties~\cite{DiCintio:2013qxa, Bose:2018oaj, Read:2015sta, Kim:2021zzw}). We include this effect by setting cored profiles above a threshold halo mass $M^\mathrm{core}_\mathrm{thres}$. We use the simulation- and observation-constrained profile from Ref.~\cite{Read:2015sta}, where the core size depends on the time over which a galaxy has formed stars, assuming that satellites stop forming stars at infall. $M^\mathrm{core}_\mathrm{thres}$ is uncertain, so in our data analysis below we set it to be a free parameter. This allows us to observationally explore the cusp-core transition.

The Milky Way gravitational field can tidally strip DM subhalos, changing their density profile (we discuss tidal disruption at the end of this Section). This may be modeled by truncating the density profile beyond the tidal radius~\cite{Baltz:2007vq, Gilman:2019nap, Gilman:2021gkj}, defined as the radius where the Milky Way tidal force balances the attractive force of the subhalo~\cite{King:1962wi}; and by modifying structural halo parameters due to tidal shocking and heating~\cite{Gnedin:1997vp, Read:2005zm, Dooley:2016ajo, Delos:2019lik, Drakos:2020ksc, Penarrubia:2010jk, Errani:2020wgn, 2022MNRAS.517.1398B}. Importantly for this study, as we demonstrate in \cref{sec:app_baryon} (\cref{fig:tidal}), only when the \emph{average mass loss} of the satellite population is $\gtrsim 99\%$ are our results affected (our population study is sensitive to \emph{average} properties). Though uncertain, the average mass loss of undisrupted Milky Way satellites is estimated to be $\sim 30\%$--$90\%$~\cite{Garrison-Kimmel:2013eoa, Simon:2019nxf, 2023ApJ...948...87W, Grand:2021fpx, Santos-Santos:2021goz}. Hence, tidal stripping is a minor contribution to our error budget and we do not include it in our main results. Physically, this is because our observables are sensitive to halo properties inside the half-light radius, usually smaller than the tidal radius; and because galaxies remain unaffected until their DM halos lose more than $90\%$ of their mass~\cite{Kim:2021zzw, 2018MNRAS.478.3879S, Errani:2021rzi}. The high concentrations of LDM halos may also make them more resilient to tidal stripping~\cite{Dooley:2016ajo, Dai:2019lud}; further work is needed to assess this.

Regarding observational abundance, not all DM halos are expected to host galaxies. Only those more massive than $10^7$--$10^8\,M_\odot$ by reionization would have gas that can cool and form stars~\cite{2012ApJ...746..109L, Koh:2016xmf, Rey:2019zsq}. We include this via the halo occupation fraction, i.e., the probability for a halo to host a galaxy, that we parametrize as a sigmoid function
\begin{equation}
\label{eq:hof}
    \operatorname{hof}(M_{200c}) = \frac{1 + \operatorname{erf}\left(\alpha^\mathrm{hof} \log_{10}\left[M_{200c}/M_0^\mathrm{hof}\right]\right)}{2} \, ,
\end{equation}
with $\operatorname{erf}$ the error function, $M_0^\mathrm{hof}$ the mass below which halos do not host galaxies, and $\alpha^\mathrm{hof}$ a parameter controlling the steepness of the transition. With simulation and semi-analytic modelling, Refs.~\cite{Barber:2013oua, Dooley:2016xkj} found $\alpha^\mathrm{hof} = 1.3$ and $M_0^\mathrm{hof} = 10^{8.4}\,M_\odot$; there are also observational limits from the Milky Way satellite luminosity function~\cite{DES:2019ltu, Jethwa:2016gra}. In our data analysis below, we set $\alpha^\mathrm{hof}$ and $M_0^\mathrm{hof}$ to be free parameters. This allows us to observationally determine the low-mass halo occupation fraction.

\begin{figure}[b]
    \centering
    \includegraphics[width=\columnwidth]{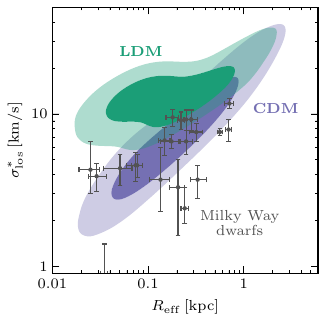}
    
    \caption{Correlation between stellar velocity dispersion and half-light radius, and data from classical and SDSS Milky Way dwarfs. Colored regions enclose 68\% and 95\% of the predicted population, for an illustrative model connecting halos to galaxies (see \cref{sec:satellites_consequences}, we later set this free). \emph{Data disfavors the high velocity dispersions of LDM models}.}
    \label{fig:data}
\end{figure}

Not all halos hosting galaxies are observable either: for a given survey, faint and distant galaxies will escape detection. We account for this by assigning to each galaxy with stellar mass below $4 \times 10^5 \, M_\odot$ (brighter galaxies correspond to classical satellites) an observation probability
\begin{equation}
    \mathcal{P}_\mathrm{obs} = \frac{\int_{V_\mathrm{obs}(L)} n(\vec{r}) \, \mathrm{d}^3r}{\int_{V_\mathrm{vir}} n(\vec{r}) \, \mathrm{d}^3r} \equiv \frac{1}{\mathcal{C}(L)}\, , \label{eq:completeness1}
\end{equation}
with $n$ the number density of satellites, $V_\mathrm{vir}$ the Milky Way virial volume, and $V_\mathrm{obs}(L)$ the volume where satellites with luminosity $L$ can be detected by the survey. $\mathcal{C}(L)$ is the luminosity-dependent completeness correction~\cite{Kim:2017iwr}. We compute luminosities assuming $M_*/M_\odot = 2 L/L_\odot$, appropriate for old stellar populations~\cite{Wolf:2009tu, Woo:2008gg}. 

\emph{A priori}, $\mathcal{P}_\mathrm{obs}$ depends both on $L$ and $R_\mathrm{eff}$~\cite{Koposov:2007ni, 2022MNRAS.516.3944M}, but the latter dependence can be absorbed through the empirical $R_\mathrm{eff}$--$M_*$ relation of observed galaxies~\cite{Walsh:2008qn}, producing an \emph{effective} $\mathcal{P}_\mathrm{obs}$ that only depends on $L$. Uncertainties on that relation do not significantly affect our error budget, see \cref{sec:app_statistics}. In short, uncertainties impact galaxies with very low observation probability, so our choices are appropriate to predict the population of SDSS-like galaxies. In addition, uncertainties can be partly absorbed in the stellar mass-halo mass relation and halo occupation fraction that we set free.

We assume that $n(\vec{r})$ does not change significantly with the solid angle, so that the selection function is separable into a radial and an angular component
\begin{align}
    \mathcal{C}(L) \equiv \mathcal{C}_\Omega \, \mathcal{C}_r \, ; \hspace{3pt}
    \mathcal{C}_\Omega \equiv \frac{4\pi}{\Omega} \, ;  \hspace{3pt}
    \mathcal{C}_r \equiv \frac{\int_0^{r_\mathrm{vir}} n(r) r^2 \, \mathrm{d}r}{\int_0^{r_\mathrm{c}(L)} n(r) r^2 \, \mathrm{d}r} \, , \label{eq:completeness2}
\end{align}
with $\Omega$ the angular coverage of the survey, $r_\mathrm{vir}$ the Milky Way virial radius, and $r_\mathrm{c}(L)$ the largest distance at which satellites of luminosity $L$ can be observed by the survey. We consider classical and SDSS satellites; for the latter $\Omega_\mathrm{SDSS} = 3.65\, \mathrm{sr}$ and $r_\mathrm{c}$ is given by~\cite{Walsh:2008qn, Koposov:2007ni}
\begin{equation}
    r_\mathrm{c}(L) = 1.5 \, \mathrm{kpc} \, \left(L/L_\odot\right)^{0.51}\,.
\end{equation}
$\mathcal{C}_\Omega$ has 19\% scatter due to anisotropy~\cite{Tollerud:2008ze}. As mentioned above, we neglect the impact of the Magellanic Clouds on anisotropy~\cite{2021ApJ...923..140G, 2022MNRAS.516.3944M}, because the SDSS footprint does not include them; and extra scatter of $\mathcal{C}_\Omega$ would only affect the total number of satellites that is not our main observable and has already a large error (see \cref{sec:app_statistics}).

The Milky Way can also tidally disrupt DM halos. This would remove galaxies from our sample and affect the radial distribution of undisrupted satellites $n(r)$, which is hence very uncertain~\cite{Kim:2017iwr, Garrison-Kimmel:2017zes, Gnedin:2004cx, Abadi:2009ve, DOnghia:2009xhq, Brooks:2012ah, Sawala:2016tlo, Samuel:2019ylk, 2020MNRAS.492.5780R, 2022ApJ...940..136P, Grand:2021fpx}. This made it a big systematic uncertainty in previous work of satellite populations~\cite{Kim:2017iwr, DES:2019ltu}. To be conservative, we are agnostic about tidal disruption and include both effects by parametrizing $n(r)$ as
\begin{equation}
    n(r) = n_\mathrm{GK17}(r) + y_\mathcal{C} [n_\mathrm{NFW}(r) - n_\mathrm{GK17}(r)] \, ,
    \label{eq:n_radial}
\end{equation}
with $y_\mathcal{C} \in [0, 1]$ a parameter that interpolates between an unstripped NFW distribution, $n_\mathrm{NFW}(r)$ --- i.e., satellites follow the DM halo distribution ---, and the strongly disrupted distribution from Ref.~\cite{Garrison-Kimmel:2017zes}, $n_\mathrm{GK17}(r)$ (named \texttt{FIRE} in Ref.~\cite{Garrison-Kimmel:2017zes}, see also the dotted line in Fig.~1 of Ref.~\cite{Kim:2017iwr}). In our analysis below, we set $y_\mathcal{C}$ free, which allows to observationally explore tidal subhalo disruption.

\subsection{Consequences}
\label{sec:satellites_consequences}

\Cref{fig:data} illustrates how the large concentrations predicted in LDM models get imprinted on galaxies in the form of high stellar velocity dispersions, which can be in tension with Milky Way satellite data. We plot the correlation between $\sigma_\mathrm{los}^*$ and $R_\mathrm{eff}$ of Milky Way satellites. We generate the theoretical expectations by sampling halo masses from the product of the subhalo mass function described in \cref{sec:consequences_DM_formalism} and the halo occupation fraction from Ref.~\cite{Dooley:2016xkj}, and sampling halo concentrations as described in \cref{sec:consequences_DM_formalism}. We use the same LDM parameters as in \cref{fig:physics,fig:hmf}, $k_\mathrm{cut}=8\,\mathrm{Mpc^{-1}}$ and $n_\mathrm{cut}=2.6$ (our analysis below excludes this). We assign $R_\mathrm{eff}$ and $\sigma_\mathrm{los}^*$ using \cref{eq:sigma_los,eq:R_eff}. $R_\mathrm{eff}$ depends on the stellar mass, that we generate using the stellar mass-halo mass relation from Ref.~\cite{Moster:2012fv} with lognormal scatter of 0.15\,dex~\cite{Moster:2012fv}. $\sigma_\mathrm{los}^*$ depends on the halo density profile: we set cored profiles for halo masses above $M_\mathrm{thres}^\mathrm{core} = 10^9 M_\odot$~\cite{Kim:2021zzw}. Finally, we weight each galaxy by its observation probability $\mathcal{P}_\mathrm{obs}$, computed with \cref{eq:completeness1} for an undisrupted NFW spatial distribution of satellites (i.e., $y_\mathcal{C}=1$ in \cref{eq:n_radial}). Data correspond to classical and SDSS satellites, as obtained from Refs.~\cite{McConnachie:2012vd, Kim:2021zzw, Kirby:2013isa} and compiled in \cref{sec:app_data}. In our analysis below, we allow the parameters of the stellar mass-halo mass relation, the halo occupation fraction, the halo mass where baryonic feedback switches the profiles from NFW to cored, and the amount of tidal subhalo disruption to vary; this significantly affects the predicted population (see \cref{sec:app_baryon}). It also increases the agreement of data with CDM, although our choices for \cref{fig:data} are within the 2$\sigma$ allowed region (see \cref{sec:app_full_results}).

As we see from \cref{fig:data}, velocity dispersion $\sigma_\mathrm{los}^*$ is correlated with half-light radius $R_\mathrm{eff}$. Galaxies with low $R_\mathrm{eff}$ are less massive (see \cref{eq:Mstar,eq:R_eff}), which leads to lower $\sigma_\mathrm{los}^*$ (see \cref{eq:sigma_los}). However, $\sigma_\mathrm{los}^*$ also depends on the central density, i.e., on the halo concentration. As in LDM models low-mass halos have higher concentrations than in CDM, LDM predicts higher $\sigma_\mathrm{los}^*$ at low $R_\mathrm{eff}$. 

These results highlight that \emph{taking into account stellar kinematics and their correlation with the half-light radius is a powerful observational probe of the DM power spectrum}. Below, we carry out a statistical analysis of classical and SDSS Milky Way satellites to quantify this.

\begin{figure*}[ht]
    \centering
    \includegraphics[width=1.01\columnwidth, valign=t]{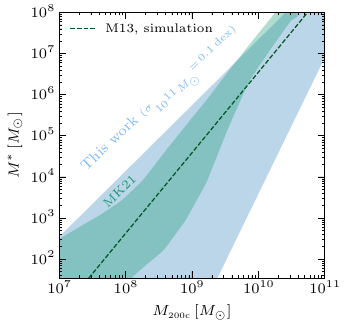} 
    \includegraphics[width=1.01\columnwidth, valign=t]{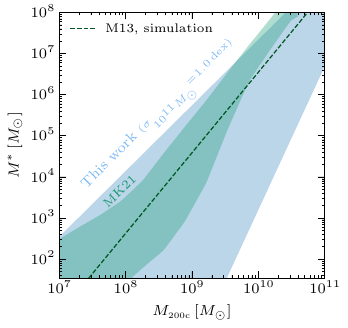}
        \vspace{-0.2cm}
    \caption{Stellar mass-halo mass relation allowed by our analysis within $2\sigma$, for different fixed scatters at high mass, together with the simulation prediction that we use in \cref{fig:data} (Moster et al.~2013, M13~\cite{Moster:2012fv}) and previous constraints (Manwadkar and Kravtsov 2021, MK21~\cite{2022MNRAS.516.3944M}). \emph{Increased scatter suppresses the median stellar mass to avoid overproducing bright galaxies. The lower mass end is degenerate with the halo occupation fraction and is hence more uncertain.}}
    \label{fig:smhm}
        \vspace{-0.3cm}
\end{figure*}

\section{Statistical analysis and results}
\label{sec:analysis}

The procedure described above allows us to predict the number of visible Milky Way satellites, as well as their line-of-sight stellar velocity dispersions $\sigma_\mathrm{los}^*$ and half-light radii $R_\mathrm{eff}$, for CDM and LDM models. In this section, we use observational data to infer the primordial DM power spectrum and the galaxy-halo connection parameters.

We fit the number of classical and SDSS Milky Way satellites, their velocity dispersions, and half-light radii; as obtained from Refs.~\cite{McConnachie:2012vd, Kim:2021zzw, Kirby:2013isa} and compiled in \cref{sec:app_data}; using an unbinned likelihood described in \cref{sec:app_statistics}. We generate theoretical predictions with \texttt{Galacticus}, for DM halo properties; and \href{https://github.com/stacykim/disSat}{\texttt{dis}}, for galaxy properties; as described in \cref{sec:consequences_DM,sec:consequences_MW}. Our free parameters are $\{k_\mathrm{cut},\,n_\mathrm{cut}\}$, that describe the DM power spectrum; \{$\beta^{M_*}$, $\sigma^{M_*}$, $\gamma^{M_*}$, $\alpha^\mathrm{hof}$, $M_0^\mathrm{hof}$, $M_\mathrm{thres}^\mathrm{core}$\}, that describe the galaxy-halo connection; and $y_\mathcal{C}$, that parametrizes tidal disruption of Milky Way subhalos. Since the anisotropy of the satellite distribution has 19\% scatter~\cite{Tollerud:2008ze}, we multiply $\mathcal{C}_\Omega$ in \cref{eq:completeness2} with a free parameter $\sigma_{\mathcal{C}_\Omega}$ and we add a Gaussian prior on  $\sigma_{\mathcal{C}_\Omega}$ centered at 1 and with a width of 0.19. $\sigma_{\mathcal{C}_\Omega}$ is our last free parameter.

We perform a frequentist profile-likelihood analysis to avoid prior dependence and Bayesian volume effects. We expect these effects because some parameters are not strongly constrained, there are non-trivial correlations, and part of the parameter space is unbounded.

\subsection{Galaxy properties}
\label{sec:results_galaxy}

Our analysis constrains the relation between galaxies and DM halos with minimal assumptions on baryonic physics \emph{and} the small-scale DM power spectrum. Below, we describe our inferences on the stellar mass-halo mass relation, the halo occupation fraction, and the amount of subhalo tidal disruption by the Milky Way. We provide the full results of the analysis in \cref{sec:app_full_results}.

\Cref{fig:smhm} shows the stellar mass-halo mass relation allowed by our analysis within $2\sigma$, compared against observational constraints using the luminosity function~\cite{2022MNRAS.516.3944M} (which includes DES and PS1 satellites on top of SDSS and classical satellites, assumes a standard CDM power spectrum, and uses a model for galaxy luminosity and halo occupation fraction), and the extrapolated predictions from the simulation of Ref.~\cite{Moster:2012fv}. Since we describe the stellar mass-halo mass relation in terms of three parameters --- the median power-law slope, the scatter at $10^{11} \, M_\odot$, and the growth of scatter at low mass, see \cref{eq:Mstar,eq:Mstar_scatter} --- and we find significant correlations (see \cref{sec:app_full_results}), we show the allowed relation for two fixed values of the scatter at $10^{11} \, M_\odot$: 0.1\,dex and 1\,dex. Other parameters, including those controlling the DM power spectrum, are allowed to vary freely.

Our median results are compatible with extrapolation from simulation~\cite{Moster:2012fv}. We see that the median power-law slope is strongly correlated with scatter: if scatter is high, the median stellar mass of low-mass halos has to be small. Otherwise, too many visible satellite galaxies would be predicted. This allows us to obtain a quite robust upper limit on the largest stellar mass of low-mass halos, consistent with previous work~\cite{2022MNRAS.516.3944M}.

The smallest stellar mass of low-mass halos, however, is worse-constrained by our analysis. This is because decreasing the stellar mass would predict less visible satellites, which is partly degenerate with increased halo occupation fraction. \emph{A priori} information on the halo occupation fraction (as in Ref.~\cite{2022MNRAS.516.3944M}) would break this degeneracy; we comment on this in the Conclusions.

We also conclude that \emph{the uncertain stellar mass-halo mass relation should not strongly affect our determination of the DM power spectrum}. The main observable for the stellar mass-halo mass relation is the number of observed satellites, whereas the power spectrum is determined mainly from the correlation between stellar velocity dispersion and half-light radius, see \cref{fig:data}. There is only a mild degeneracy because by making low-mass halos brighter, the average velocity dispersion in LDM models can be somewhat reduced. The correlations in our full analysis in \cref{sec:app_full_results} confirm this intuition.

\begin{figure}[t]
    \centering
    \includegraphics[width=\columnwidth]{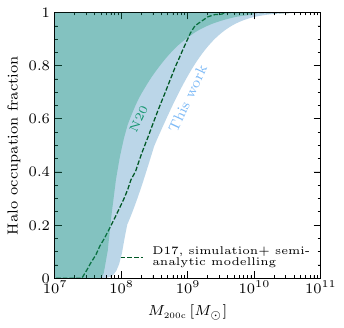}
    \vspace{-0.3cm}
    \caption{Halo occupation fraction allowed by our analysis within $2\sigma$, the prediction used in \cref{fig:data} (Dooley et al.~2017, D17~\cite{Dooley:2016xkj}), and previous constraints (Nadler et al.~2020, N20~\cite{DES:2019ltu}). Regions within the arrows are allowed. \emph{Our analysis informs on halo occupation at masses above $\sim 10^8 \, M_\odot$}.}
    \vspace{-0.3cm}
    \label{fig:hof}
\end{figure}

\Cref{fig:hof} shows the halo occupation fraction allowed by our analysis within $2\sigma$, together with existing constraints using the luminosity function~\cite{DES:2019ltu} (using DES and PS1 dwarfs and assuming a standard CDM power spectrum), and the relation inferred in Refs.~\cite{Barber:2013oua, Dooley:2016xkj} by combining simulation with semi-analytic modelling (see also Refs.~\cite{Jethwa:2016gra, 2021ApJ...923...35M}). All the region within the arrows is allowed. Other parameters, including those controlling the DM power spectrum, are allowed to vary freely.

We observe that data requires halos with masses between $10^8 M_\odot$ and $10^{10} M_\odot$ to host galaxies. This is consistent with the results from recent simulations and the luminosity function: if these halos did not host galaxies, there would be a ``too many satellites'' problem. Our analysis cannot robustly decide if lower-mass halos have a non-zero probability of not hosting galaxies (i.e., if their halo occupation fraction is not 0), as this is degenerate with a modified stellar mass-halo mass relation: very low-mass halos can host galaxies if they are very dim.

In our analysis, the halo occupation fraction is mildly degenerate with the DM power spectrum, as by populating low-mass halos the average velocity dispersion in LDM models can be reduced. However, the additional handle from the total number of galaxies together with the different correlation between $\sigma_\mathrm{los}^*$ and $R_\mathrm{eff}$ in LDM and CDM models (see \cref{fig:data}) alleviate this effect.

Regarding tidal subhalo disruption and the radial satellite distribution in \cref{eq:completeness1,eq:completeness2,eq:n_radial}, we find $y_\mathcal{C} \gtrsim 0.4$ at $1\sigma$ (see \cref{sec:app_full_results}). I.e., our analysis disfavors strong tidal disruption found in simulation~\cite{Garrison-Kimmel:2017zes}, as that would produce a ``too many satellites'' problem~\cite{Kim:2017iwr}.

\subsection{Dark Matter properties}
\label{sec:results_DM}
We now turn to the original purpose of this paper: \emph{what do Milky Way satellite properties tell us about the DM power spectrum at the smallest scales}? 

\begin{figure}[b]
    \centering
    \vspace{-0.2cm}
    \includegraphics[width=0.95\columnwidth]{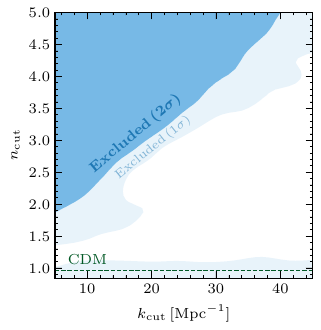}
    \vspace{-0.5cm}
    \caption{Excluded values of LDM parameter space in our analysis. \emph{Dwarf galaxy velocity dispersions and sizes probe dark matter substructure for 5\,Mpc$^{-1} \lesssim k \lesssim$ 40\,Mpc$^{-1}$.}}
    \label{fig:LDM_constraints}
\end{figure}

\begin{figure*}[t]
    \centering
    \includegraphics[width=\textwidth]{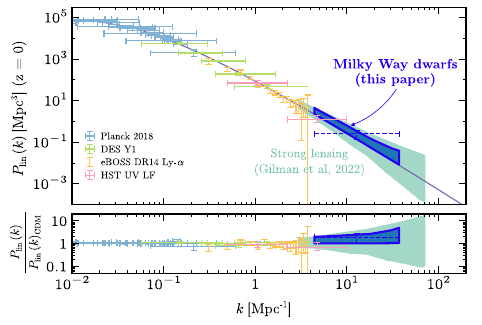}  
    
    \caption{1$\sigma$ allowed linear matter power spectrum in our analysis (shaded), its geometric mean (dashed), and previous data~\cite{Planck:2018nkj, DES:2017qwj, SDSS:2017yll, Chabanier:2019eai, Sabti:2021unj, Gilman:2021gkj}. \emph{Dwarf galaxy velocity dispersion and size data is a unique, sharp probe of small-scale dark matter properties.}}
    \label{fig:constraints}
\end{figure*}

\Cref{fig:LDM_constraints} shows that the DM power spectrum cannot be strongly enhanced at dwarf galaxy scales. The allowed parameter space is unbounded because very large $k_\mathrm{cut}$ is indistinguishable from scale invariance, but our analysis limits how large enhancements can be between comoving scales 5\,Mpc$^{-1} \lesssim k \lesssim$ 40\,Mpc$^{-1}$ (as large values of the power-spectrum slope $n_\mathrm{cut}$ are excluded). We show the $1\sigma$ and $2\sigma$ excluded values of $k_\mathrm{cut}$ and $n_\mathrm{cut}$, allowing other parameters to vary freely. We find a slight $\sim 1\sigma$ preference for enhanced power, which is due to the slight ``turn up'' of velocity dispersions at low $R_\mathrm{eff}$ in \cref{fig:data}. This effect, however, is not statistically significant.

The results are intuitive: from \cref{fig:data}, data approximately follow the CDM correlation between $\sigma_\mathrm{los}^*$ and $R_\mathrm{eff}$, with no clear sign of a break characteristic of LDM. This favors similar concentrations for all Milky Way satellites, i.e., similar formation times. As explained in \cref{sec:consequences_DM_consequences}, this favors close-to-scale-invariant power spectra.

Our analysis loses sensitivity for $k_\mathrm{cut} \lesssim 5 \, \mathrm{Mpc^{-1}}$ and ${k_\mathrm{cut} \gtrsim 40 \, \mathrm{Mpc^{-1}}}$. This is intuitive: using \cref{eq:window}, smaller $k$ correspond to halo masses $M_{200 c} \gtrsim 10^{11} M_\odot$, above the mass of a Milky Way satellite galaxy. Larger $k$ correspond to halo masses $M_{200 c} \lesssim 10^{8} M_\odot$. The dwarfs with lowest half-light radii that we consider (Segue-I, Segue-II and Willman-I)  have ${M_* \simeq 2\text{--}5 \times 10^{2} M_\odot}$, i.e., halo masses in that ballpark (see \cref{fig:smhm}).

Our determination of the power spectrum is robust against baryonic effects and different galaxy-halo connection models. The strongest degeneracy we find is with the halo mass above which baryonic feedback makes DM density profiles cored, $M_\mathrm{thres}^\mathrm{core}$. By making low-mass halos cored, their velocity dispersion gets reduced, and LDM predictions resemble more those of CDM. However, a CDM-like correlation between $\sigma_\mathrm{los}^*$ and $\mathrm{R}_\mathrm{eff}$ (see \cref{fig:data}) is never fully mimicked. Data can tell apart LDM and CDM even if LDM halos are cored at all halo masses.

\Cref{fig:constraints} shows that dwarf galaxy velocities and sizes determine the DM power spectrum at small scales with unprecedented precision. We show existing measurements from Planck~\cite{Planck:2018nkj}, DES~\cite{DES:2017qwj}, eBOSS Lyman-$\alpha$~\cite{SDSS:2017yll, Chabanier:2019eai}, the UV luminosity function as measured by the Hubble Space Telescope~\cite{Sabti:2021unj}, and strong gravitational lensing~\cite{Gilman:2021gkj}; the first three measurements have been obtained using the \href{https://github.com/marius311/mpk\_compilation}{\texttt{mpk\_compilation}} code~\cite{Chabanier:2019eai}. Since our parametrization in \cref{eq:powerSpec} only allows for enhanced power spectra, for consistency we have ensured that the analysis does not prefer suppressed power spectra. To do so, we parametrize suppressed power with $n_\mathrm{cut} < n_s$ as $\mathcal{P}_\mathcal{R}(k) \propto k^{n_s-1} \left/ \left[1 + \left(k/k_\mathrm{cut}\right)^{n_s - n_\mathrm{cut}}\right]\right.$ and repeat our analysis extending the range to $n_\mathrm{cut}<1$. That expression smoothly interpolates from $\mathcal{P}_\mathcal{R} \propto k^{n_s-1}$ at $k \ll k_\mathrm{cut}$ to $\mathcal{P}_\mathcal{R} \propto k^{n_\mathrm{cut}-1}$ at $k \gg k_\mathrm{cut}$. To aid the connection with $k_\mathrm{cut}$, we use units of $\mathrm{Mpc^{-1}}$ instead of $h\,\mathrm{Mpc^{-1}}$.

The dark shaded area in \cref{fig:constraints} encloses all power spectra in our 1$\sigma$ allowed region for \{$k_\mathrm{cut}$, $n_\mathrm{cut}$\}, and the dashed data point is its geometric mean. Quantitatively,
\begin{equation}
    \frac{P_\mathrm{lin}}{P_\mathrm{lin}^\mathrm{CDM}} = 1.81 \pm 0.80 \, ,
\end{equation}
for $4 \, \mathrm{Mpc^{-1}} < k < 37 \, \mathrm{Mpc^{-1}}$. The range in $k$ is where our analysis provides significant constraints. For other values of $k$, our uncertainties increase rapidly.

Our results are somewhat stronger than, and complementary to, the strong lensing determination in Ref.~\cite{Gilman:2021gkj}, which has different model assumptions and uncertainties. 

Overall, our results add a high-precision, small-scale handle to the global picture of the primordial DM power spectrum being almost-scale-invariant across 6 orders of magnitude in wavenumber. We do so by exploiting sizes and kinematics of one of the smallest DM-dominated class of systems we know of: Milky Way satellite galaxies.

\section{Conclusions and future directions}
\label{sec:conclusions}

The properties of DM at small scales remain uncertain. In this paper, we have shown that the correlation between dwarf galaxy internal velocities and sizes is a powerful probe to explore them. Our constraining power mostly comes from probing the concentration-mass relation of DM halos, in contrast to previous work focusing on halo abundances~\cite{DES:2020fxi, DES:2022doi, Tollerud:2008ze, Horiuchi:2013noa, kennedy2014, Kim:2017iwr, Jethwa:2016gra, DES:2020fxi, dekker2022, DES:2022doi}.

We focus on models that enhance the primordial matter power spectrum at small scales, that we denote as ``Lumpy Dark Matter'' (LDM). This prediction is natural if DM is produced out of quantum fluctuations during inflation, which does not require non-gravitational interactions with baryons~\cite{Graham:2015rva, Alonso-Alvarez:2018tus, Tenkanen:2019aij}. We utilize semi-analytic techniques and scaling relations, inspired by physics models and calibrated against simulations, to compute DM halo properties. We are agnostic about many aspects of the galaxy-halo connection, allowing us to marginalize over poorly understood baryonic effects, and carry out a joint analysis of internal velocities, sizes, and total abundance of classical and SDSS Milky Way satellites.

Regardless of the DM power spectrum, we constrain the stellar mass-halo mass relation, the halo occupation fraction, and tidal subhalo disruption. We find that halos with masses of $10^8 M_\odot$ and $10^9 M_\odot$ must host galaxies with stellar masses below $10^4 M_\odot$ and $10^6 M_\odot$, respectively; and our analysis disfavors strong tidal disruption. This is mostly inferred from the total satellite abundance, implying \emph{little correlation with our determination of DM properties} that relies mostly on the correlation between stellar velocity dispersion and half-light radius.

We obtain a leading-precision determination of the DM power spectrum on sub-galactic scales $4 \, \mathrm{Mpc^{-1}} < k < 37 \, \mathrm{Mpc^{-1}}$. Milky Way satellite properties imply that, at those scales, the DM power spectrum cannot deviate from scale invariance by more than a factor $\sim 2.5$ at $1\sigma$. 

Our methodology can be easily extended in further work. On the DM side, many scenarios that modify the small-scale DM properties have been explored in the literature. Examples include Self-Interacting Dark Matter (SIDM)~\cite{Tulin:2017ara}, Warm Dark Matter (WDM)~\cite{Dodelson:1993je, Ellis:1983ew}, Fuzzy Dark Matter (FDM)~\cite{Hu:2000ke}, or a modified cosmological history~\cite{StenDelos:2019xdk, Delos:2023vfv}. They are theoretically and observationally motivated, and in many of them the power spectrum at small scales is suppressed. Their consequences for the correlation between velocity dispersions and half-light radii are unexplored. The fact that our analysis finds a slight preference for enhanced power spectra forecasts increased sensitivity to such models.

On the modeling side, in this paper we have been agnostic about the galaxy-halo connection. However, there are physically motivated models that predict the stellar mass-halo mass relation and the halo occupation fraction, as a function of halo concentration and assembly history~\cite{Rey:2019zsq, 2022MNRAS.514.2667K, Edge_DarkLight, Kravtsov:2023oxa}. Adding that information would improve the sensitivity to DM properties. For instance, higher concentration and earlier halo formation in LDM models may increase the stellar mass and the probability to host a galaxy~\cite{Hirano:2015wla} (as halos would be more massive before reionization and may form stars more efficiently), contradicting observations. Our data analysis methodology can also be used to test galaxy formation models. As mentioned in the Introduction, future work can also improve on the results presented here by running LDM simulations of halo and galaxy formation and evolution, that could be tailored to the Milky Way assembly history and include the associated uncertainties.

On the observational side, there is plenty of room for increased precision. We have only included classical and SDSS satellites (whose observations could be improved~\cite{Simon:2019nxf}), but our analysis can be extended to include Milky Way satellites discovered by the Dark Energy Survey, for which we are starting to get kinematic data (see, e.g., Refs~\cite{2018ApJ...852...68C, 2018ApJ...857...70C, 2018arXiv180902259J, 2018ApJ...863...25M, DES:2020szo}) and whose completeness correction is understood~\cite{DES:2019vzn}; Milky Way satellites discovered by DELVE~\cite{DELVE:2019xvr, DELVE:2021rcc, DELVE:2021lds, DELVE:2022ijm, DELVE:2022jls}; M31 satellites, to avoid the possibility of the Milky Way being an outlier; or even field dwarfs, unaffected by host galaxy effects but whose completeness correction and DM-driven velocity dispersion is harder to model. In the future, the Rubin Observatory should discover many dwarfs, within and outside the Milky Way, and inform on their spatial distribution~\cite{LSSTScience:2009jmu, Mutlu-Pakdil:2021crk}. This would also quantify the amount of tidal disruption, which may be suppressed in LDM models as discussed above.

The satellite luminosity function can also be added to our dataset. This would better constrain the stellar mass-halo mass relation and the halo occupation fraction. Abundances, luminosities, sizes, and kinematics are the galaxy observables that most directly relate to DM halo features. Analyzing them together would provide a more complete picture of DM properties. 

Finally, LDM models have a rich phenomenology that can be explored with other observables. As \cref{fig:physics} shows, enhanced power spectra dramatically increase the formation redshift of low-mass halos, by breaking a fortuitous cancellation in the density variance of scale-invariant power spectra. This may produce stars early~\cite{Hirano:2015wla} and affect reionization, which can be explored with Cosmic Microwave Background~\cite{Planck:2016mks, Planck:2018vyg, Castellano:2022knz} and 21\,cm observations~\cite{Furlanetto:2015apc}. JWST observations are also providing a unique window to the high-redshift Universe and the formation of the first galaxies~\cite{2022ApJ...940L..14N, 2022ApJ...938L..15C, 2023ApJ...948L..14C, 2023MNRAS.518.4755A, 2023MNRAS.519.1201A, 2023MNRAS.519.3691D, 2023MNRAS.520.4554D, 2023ApJS..265....5H, 2023MNRAS.523.1009B, 2023MNRAS.523.1036B, 2022ApJ...938..144M, 2023arXiv230406658H, Dayal:2023nwi}. Precisely determining the UV luminosity function can be a sharp probe of the DM power spectrum, as shown by existing work with HST data~\cite{Sabti:2021unj, Menci:2016eui} and first JWST results~\cite{Sabti:2023xwo, Parashari:2023cui, Hirano:2023auh}.

Evidence for an unknown Dark Matter, five times more abundant than regular matter, first appeared in galaxy observations~\cite{Zwicky:1933gu, Zwicky:1937zza}. This was confirmed to high accuracy observing the largest scales of our Universe~\cite{WMAP:2008lyn, Planck:2018vyg}. By studying back small-scale Dark Matter-dominated systems with increased precision, we are now on our way to building a coherent picture of the properties of the second most abundant component of the Universe.

\begin{acknowledgments}

We thank Andrew Benson for helpful comments, a careful reading of the manuscript, and collaboration in early stages of the project. We are grateful for helpful comments from John Beacom, Hector Cruz, Nicole Gountains, Zavier Kamath, and especially Xiaolong Du, Daniel Gillman, Jordi Miralda-Escude, Ethan Nadler, Nashwan Sabti, Chun-Hao To, and Katy Rodriguez Wimberly.  

AHGP's work is supported in part by NSF Grant No.~AST-2008110 and the NASA Astrophysics Theory Program, under grant 80NSSC18K1014. IE acknowledges support from Basque Government (IT1628-22); the PID2021-123703NB-C21 grant funded by MCIN/\allowbreak AEI/10.13039/501100011033/ and by ERDF ``A way of making Europe''; and the PID2022-136510NB-C33 grant funded by MCIN/AEI. Part of this work used the Solaris cluster, acquired through the Basque Government IT1628-22 grant.

\end{acknowledgments}

\bibliographystyle{JHEP.bst}
\bibliography{References}

\providecommand{\href}[2]{#2}\begingroup\raggedright\begin{thebibliography}{100}

\bibitem{Planck:2018vyg}
{\bf Planck} {\bf Collaboration}, N.~Aghanim {\em et~al.}, {\it {Planck 2018
  results. VI. Cosmological parameters}},  {\em Astron. Astrophys.} {\bf 641}
  (2020) A6, [\href{http://www.arxiv.org/abs/1807.06209}{{\tt 1807.06209}}].
  [Erratum: Astron.Astrophys. 652, C4 (2021)].

\bibitem{DES:2021wwk}
{\bf DES} {\bf Collaboration}, T.~M.~C. Abbott {\em et~al.}, {\it {Dark Energy
  Survey Year 3 results: Cosmological constraints from galaxy clustering and
  weak lensing}},  {\em Phys. Rev. D} {\bf 105} (2022), no.~2 023520,
  [\href{http://www.arxiv.org/abs/2105.13549}{{\tt 2105.13549}}].

\bibitem{SDSS:2017yll}
{\bf SDSS} {\bf Collaboration}, M.~R. Blanton {\em et~al.}, {\it {Sloan Digital
  Sky Survey IV: Mapping the Milky Way, Nearby Galaxies and the Distant
  Universe}},  {\em Astron. J.} {\bf 154} (2017), no.~1 28,
  [\href{http://www.arxiv.org/abs/1703.00052}{{\tt 1703.00052}}].

\bibitem{Bullock:2017xww}
J.~S. Bullock and M.~Boylan-Kolchin, {\it {Small-Scale Challenges to the
  $\Lambda$CDM Paradigm}},  {\em Ann. Rev. Astron. Astrophys.} {\bf 55} (2017)
  343--387, [\href{http://www.arxiv.org/abs/1707.04256}{{\tt 1707.04256}}].

\bibitem{Buckley:2017ijx}
M.~R. Buckley and A.~H.~G. Peter, {\it {Gravitational probes of dark matter
  physics}},  {\em Phys. Rept.} {\bf 761} (2018) 1--60,
  [\href{http://www.arxiv.org/abs/1712.06615}{{\tt 1712.06615}}].

\bibitem{crnojevic2021}
D.~{Crnojevi{\'c}} and B.~{Mutlu-Pakdil}, {\it {Dwarf galaxies yesterday, now
  and tomorrow}},  {\em Nature Astronomy} {\bf 5} (Dec., 2021) 1191--1194.

\bibitem{sales2022}
L.~V. {Sales}, A.~{Wetzel}, and A.~{Fattahi}, {\it {Baryonic solutions and
  challenges for cosmological models of dwarf galaxies}},  {\em Nature
  Astronomy} {\bf 6} (June, 2022) 897--910,
  [\href{http://www.arxiv.org/abs/2206.05295}{{\tt 2206.05295}}].

\bibitem{Tulin:2017ara}
S.~Tulin and H.-B. Yu, {\it {Dark Matter Self-interactions and Small Scale
  Structure}},  {\em Phys. Rept.} {\bf 730} (2018) 1--57,
  [\href{http://www.arxiv.org/abs/1705.02358}{{\tt 1705.02358}}].

\bibitem{Dodelson:1993je}
S.~Dodelson and L.~M. Widrow, {\it {Sterile-neutrinos as dark matter}},  {\em
  Phys. Rev. Lett.} {\bf 72} (1994) 17--20,
  [\href{http://www.arxiv.org/abs/hep-ph/9303287}{{\tt hep-ph/9303287}}].

\bibitem{Ellis:1983ew}
J.~R. Ellis, J.~S. Hagelin, D.~V. Nanopoulos, K.~A. Olive, and M.~Srednicki,
  {\it {Supersymmetric Relics from the Big Bang}},  {\em Nucl. Phys. B} {\bf
  238} (1984) 453--476.

\bibitem{Hu:2000ke}
W.~Hu, R.~Barkana, and A.~Gruzinov, {\it {Cold and fuzzy dark matter}},  {\em
  Phys. Rev. Lett.} {\bf 85} (2000) 1158--1161,
  [\href{http://www.arxiv.org/abs/astro-ph/0003365}{{\tt astro-ph/0003365}}].

\bibitem{brooks2019}
A.~M. {Brooks}, {\it {Understanding Dwarf Galaxies in Order to Understand Dark
  Matter}},  in {\em Illuminating Dark Matter} (R.~{Essig}, J.~{Feng}, and
  K.~{Zurek}, eds.), vol.~56 of {\em Astrophysics and Space Science
  Proceedings}, p.~19, Jan., 2019.
\newblock \href{http://www.arxiv.org/abs/1812.00044}{{\tt 1812.00044}}.

\bibitem{Tollerud:2008ze}
E.~J. Tollerud, J.~S. Bullock, L.~E. Strigari, and B.~Willman, {\it {Hundreds
  of Milky Way Satellites? Luminosity Bias in the Satellite Luminosity
  Function}},  {\em Astrophys. J.} {\bf 688} (2008) 277--289,
  [\href{http://www.arxiv.org/abs/0806.4381}{{\tt 0806.4381}}].

\bibitem{Horiuchi:2013noa}
S.~Horiuchi, P.~J. Humphrey, J.~Onorbe, K.~N. Abazajian, M.~Kaplinghat, and
  S.~Garrison-Kimmel, {\it {Sterile neutrino dark matter bounds from galaxies
  of the Local Group}},  {\em Phys. Rev. D} {\bf 89} (2014), no.~2 025017,
  [\href{http://www.arxiv.org/abs/1311.0282}{{\tt 1311.0282}}].

\bibitem{kennedy2014}
R.~{Kennedy}, C.~{Frenk}, S.~{Cole}, and A.~{Benson}, {\it {Constraining the
  warm dark matter particle mass with Milky Way satellites}},  {\em Mon. Not.
  Roy. Astron. Soc.} {\bf 442} (Aug., 2014) 2487--2495,
  [\href{http://www.arxiv.org/abs/1310.7739}{{\tt 1310.7739}}].

\bibitem{Kim:2017iwr}
S.~Y. Kim, A.~H.~G. Peter, and J.~R. Hargis, {\it {Missing Satellites Problem:
  Completeness Corrections to the Number of Satellite Galaxies in the Milky Way
  are Consistent with Cold Dark Matter Predictions}},  {\em Phys. Rev. Lett.}
  {\bf 121} (2018), no.~21 211302,
  [\href{http://www.arxiv.org/abs/1711.06267}{{\tt 1711.06267}}].

\bibitem{Jethwa:2016gra}
P.~Jethwa, D.~Erkal, and V.~Belokurov, {\it {The upper bound on the lowest mass
  halo}},  {\em Mon. Not. Roy. Astron. Soc.} {\bf 473} (2018), no.~2
  2060--2083, [\href{http://www.arxiv.org/abs/1612.07834}{{\tt 1612.07834}}].

\bibitem{DES:2020fxi}
{\bf DES} {\bf Collaboration}, E.~O. Nadler {\em et~al.}, {\it {Milky Way
  Satellite Census. III. Constraints on Dark Matter Properties from
  Observations of Milky Way Satellite Galaxies}},  {\em Phys. Rev. Lett.} {\bf
  126} (2021) 091101, [\href{http://www.arxiv.org/abs/2008.00022}{{\tt
  2008.00022}}].

\bibitem{dekker2022}
A.~{Dekker}, S.~{Ando}, C.~A. {Correa}, and K.~C.~Y. {Ng}, {\it {Warm dark
  matter constraints using Milky Way satellite observations and subhalo
  evolution modeling}},  {\em \prd} {\bf 106} (Dec., 2022) 123026,
  [\href{http://www.arxiv.org/abs/2111.13137}{{\tt 2111.13137}}].

\bibitem{DES:2022doi}
{\bf DES} {\bf Collaboration}, S.~Mau {\em et~al.}, {\it {Milky Way Satellite
  Census. IV. Constraints on Decaying Dark Matter from Observations of Milky
  Way Satellite Galaxies}},  {\em Astrophys. J.} {\bf 932} (2022), no.~2 128,
  [\href{http://www.arxiv.org/abs/2201.11740}{{\tt 2201.11740}}].

\bibitem{Akita:2023yga}
K.~Akita and S.~Ando, {\it {Constraints on dark matter-neutrino scattering from
  the Milky-Way satellites and subhalo modeling for dark acoustic
  oscillations}},  \href{http://www.arxiv.org/abs/2305.01913}{{\tt
  2305.01913}}.

\bibitem{Kim:2021zzw}
S.~Y. Kim and A.~H.~G. Peter, {\it {The Milky Way satellite velocity function
  is a sharp probe of small-scale structure problems}},
  \href{http://www.arxiv.org/abs/2106.09050}{{\tt 2106.09050}}.

\bibitem{Subramanian:2015lua}
K.~Subramanian, {\it {The origin, evolution and signatures of primordial
  magnetic fields}},  {\em Rept. Prog. Phys.} {\bf 79} (2016), no.~7 076901,
  [\href{http://www.arxiv.org/abs/1504.02311}{{\tt 1504.02311}}].

\bibitem{Achucarro:2022qrl}
A.~Ach\'ucarro {\em et~al.}, {\it {Inflation: Theory and Observations}},
  \href{http://www.arxiv.org/abs/2203.08128}{{\tt 2203.08128}}.

\bibitem{Graham:2015rva}
P.~W. Graham, J.~Mardon, and S.~Rajendran, {\it {Vector Dark Matter from
  Inflationary Fluctuations}},  {\em Phys. Rev. D} {\bf 93} (2016), no.~10
  103520, [\href{http://www.arxiv.org/abs/1504.02102}{{\tt 1504.02102}}].

\bibitem{Alonso-Alvarez:2018tus}
G.~Alonso-\'Alvarez and J.~Jaeckel, {\it {Lightish but clumpy: scalar dark
  matter from inflationary fluctuations}},  {\em JCAP} {\bf 10} (2018) 022,
  [\href{http://www.arxiv.org/abs/1807.09785}{{\tt 1807.09785}}].

\bibitem{Tenkanen:2019aij}
T.~Tenkanen, {\it {Dark matter from scalar field fluctuations}},  {\em Phys.
  Rev. Lett.} {\bf 123} (2019), no.~6 061302,
  [\href{http://www.arxiv.org/abs/1905.01214}{{\tt 1905.01214}}].

\bibitem{Zentner:2002xt}
A.~R. Zentner and J.~S. Bullock, {\it {Inflation, cold dark matter, and the
  central density problem}},  {\em Phys. Rev. D} {\bf 66} (2002) 043003,
  [\href{http://www.arxiv.org/abs/astro-ph/0205216}{{\tt astro-ph/0205216}}].

\bibitem{Zentner:2003yd}
A.~R. Zentner and J.~S. Bullock, {\it {Halo substructure and the power
  spectrum}},  {\em Astrophys. J.} {\bf 598} (2003) 49,
  [\href{http://www.arxiv.org/abs/astro-ph/0304292}{{\tt astro-ph/0304292}}].

\bibitem{Gilman:2021gkj}
D.~Gilman, A.~Benson, J.~Bovy, S.~Birrer, T.~Treu, and A.~Nierenberg, {\it {The
  primordial matter power spectrum on sub-galactic scales}},  {\em Mon. Not.
  Roy. Astron. Soc.} {\bf 512} (2022), no.~3 3163--3188,
  [\href{http://www.arxiv.org/abs/2112.03293}{{\tt 2112.03293}}].

\bibitem{Kamionkowski:1999vp}
M.~Kamionkowski and A.~R. Liddle, {\it {The Dearth of halo dwarf galaxies: Is
  there power on short scales?}},  {\em Phys. Rev. Lett.} {\bf 84} (2000)
  4525--4528, [\href{http://www.arxiv.org/abs/astro-ph/9911103}{{\tt
  astro-ph/9911103}}].

\bibitem{Sabti:2021unj}
N.~Sabti, J.~B. Mu\~noz, and D.~Blas, {\it {New Roads to the Small-scale
  Universe: Measurements of the Clustering of Matter with the High-redshift UV
  Galaxy Luminosity Function}},  {\em Astrophys. J. Lett.} {\bf 928} (2022),
  no.~2 L20, [\href{http://www.arxiv.org/abs/2110.13161}{{\tt 2110.13161}}].

\bibitem{Dalal:2001fq}
N.~Dalal and C.~S. Kochanek, {\it {Direct detection of CDM substructure}},
  {\em Astrophys. J.} {\bf 572} (2002) 25--33,
  [\href{http://www.arxiv.org/abs/astro-ph/0111456}{{\tt astro-ph/0111456}}].

\bibitem{2019MNRAS.488.4585G}
A.~S. {Graus}, J.~S. {Bullock}, T.~{Kelley}, M.~{Boylan-Kolchin},
  S.~{Garrison-Kimmel}, and Y.~{Qi}, {\it {How low does it go? Too few Galactic
  satellites with standard reionization quenching}},  {\em {Mon. Not. Roy.
  Astron. Soc.}} {\bf 488} (Oct., 2019) 4585--4595,
  [\href{http://www.arxiv.org/abs/1808.03654}{{\tt 1808.03654}}].

\bibitem{Benson:2010kx}
A.~J. Benson, {\it {Galacticus: A Semi-Analytic Model of Galaxy Formation}},
  {\em New Astron.} {\bf 17} (2012) 175--197,
  [\href{http://www.arxiv.org/abs/1008.1786}{{\tt 1008.1786}}].

\bibitem{Rocha:2011aa}
M.~Rocha, A.~H.~G. Peter, and J.~S. Bullock, {\it {Infall Times for Milky Way
  Satellites From Their Present-Day Kinematics}},  {\em Mon. Not. Roy. Astron.
  Soc.} {\bf 425} (2012) 231, [\href{http://www.arxiv.org/abs/1110.0464}{{\tt
  1110.0464}}].

\bibitem{Dooley:2016xkj}
G.~A. Dooley, A.~H.~G. Peter, T.~Yang, B.~Willman, B.~F. Griffen, and
  A.~Frebel, {\it {An observer's guide to the (Local Group) dwarf galaxies:
  predictions for their own dwarf satellite populations}},  {\em Mon. Not. Roy.
  Astron. Soc.} {\bf 471} (2017), no.~4 4894--4909,
  [\href{http://www.arxiv.org/abs/1610.00708}{{\tt 1610.00708}}].

\bibitem{2019arXiv190604180F}
S.~P. {Fillingham}, M.~C. {Cooper}, T.~{Kelley}, M.~K. {Rodriguez Wimberly},
  M.~{Boylan-Kolchin}, J.~S. {Bullock}, S.~{Garrison-Kimmel}, M.~S.
  {Pawlowski}, and C.~{Wheeler}, {\it {Characterizing the Infall Times and
  Quenching Timescales of Milky Way Satellites with $Gaia$ Proper Motions}},
  {\em arXiv e-prints} (June, 2019) arXiv:1906.04180,
  [\href{http://www.arxiv.org/abs/1906.04180}{{\tt 1906.04180}}].

\bibitem{Cautun:2019eaf}
M.~Cautun, A.~Benitez-Llambay, A.~J. Deason, C.~S. Frenk, A.~Fattahi, F.~A.
  G\'omez, R.~J.~J. Grand, K.~A. Oman, J.~F. Navarro, and C.~M. Simpson, {\it
  {The Milky Way total mass profile as inferred from Gaia DR2}},  {\em Mon.
  Not. Roy. Astron. Soc.} {\bf 494} (2020), no.~3 4291--4313,
  [\href{http://www.arxiv.org/abs/1911.04557}{{\tt 1911.04557}}].

\bibitem{Press:1973iz}
W.~H. Press and P.~Schechter, {\it {Formation of galaxies and clusters of
  galaxies by selfsimilar gravitational condensation}},  {\em Astrophys. J.}
  {\bf 187} (1974) 425--438.

\bibitem{Bond:1990iw}
J.~R. Bond, S.~Cole, G.~Efstathiou, and N.~Kaiser, {\it {Excursion set mass
  functions for hierarchical Gaussian fluctuations}},  {\em Astrophys. J.} {\bf
  379} (1991) 440.

\bibitem{Tinker:2008ff}
J.~L. Tinker, A.~V. Kravtsov, A.~Klypin, K.~Abazajian, M.~S. Warren, G.~Yepes,
  S.~Gottlober, and D.~E. Holz, {\it {Toward a halo mass function for precision
  cosmology: The Limits of universality}},  {\em Astrophys. J.} {\bf 688}
  (2008) 709--728, [\href{http://www.arxiv.org/abs/0803.2706}{{\tt
  0803.2706}}].

\bibitem{Benson:2012su}
A.~J. Benson, A.~Farahi, S.~Cole, L.~A. Moustakas, A.~Jenkins, M.~Lovell,
  R.~Kennedy, J.~Helly, and C.~Frenk, {\it {Dark Matter Halo Merger Histories
  Beyond Cold Dark Matter: I - Methods and Application to Warm Dark Matter}},
  {\em Mon. Not. Roy. Astron. Soc.} {\bf 428} (2013) 1774,
  [\href{http://www.arxiv.org/abs/1209.3018}{{\tt 1209.3018}}].

\bibitem{Leo:2018odn}
M.~Leo, C.~M. Baugh, B.~Li, and S.~Pascoli, {\it {A new smooth-$k$ space filter
  approach to calculate halo abundances}},  {\em JCAP} {\bf 04} (2018) 010,
  [\href{http://www.arxiv.org/abs/1801.02547}{{\tt 1801.02547}}].

\bibitem{Sheth:2001dp}
R.~K. Sheth and G.~Tormen, {\it {An Excursion Set Model of Hierarchical
  Clustering : Ellipsoidal Collapse and the Moving Barrier}},  {\em Mon. Not.
  Roy. Astron. Soc.} {\bf 329} (2002) 61,
  [\href{http://www.arxiv.org/abs/astro-ph/0105113}{{\tt astro-ph/0105113}}].

\bibitem{Schneider:2013ria}
A.~Schneider, R.~E. Smith, and D.~Reed, {\it {Halo Mass Function and the Free
  Streaming Scale}},  {\em Mon. Not. Roy. Astron. Soc.} {\bf 433} (2013) 1573,
  [\href{http://www.arxiv.org/abs/1303.0839}{{\tt 1303.0839}}].

\bibitem{Hirano:2015wla}
S.~Hirano, N.~Zhu, N.~Yoshida, D.~Spergel, and H.~W. Yorke, {\it {Early
  structure formation from primordial density fluctuations with a blue-tilted
  power spectrum}},  {\em Astrophys. J.} {\bf 814} (2015), no.~1 18,
  [\href{http://www.arxiv.org/abs/1504.05186}{{\tt 1504.05186}}].

\bibitem{Hirano:2023auh}
S.~Hirano and N.~Yoshida, {\it {Early Structure Formation from Primordial
  Density Fluctuations with a Blue, Tilted Power Spectrum -- II. High-Redshift
  Galaxies}},  \href{http://www.arxiv.org/abs/2306.11993}{{\tt 2306.11993}}.

\bibitem{Bohr:2021bdm}
S.~Bohr, J.~Zavala, F.-Y. Cyr-Racine, and M.~Vogelsberger, {\it {The halo mass
  function and inner structure of ETHOS haloes at high redshift}},  {\em Mon.
  Not. Roy. Astron. Soc.} {\bf 506} (2021), no.~1 128--138,
  [\href{http://www.arxiv.org/abs/2101.08790}{{\tt 2101.08790}}].

\bibitem{2019MNRAS.485.5010B}
A.~J. {Benson}, A.~{Ludlow}, and S.~{Cole}, {\it {Halo concentrations from
  extended Press-Schechter merger histories}},  {\em Mon. Not. Roy. Astron.
  Soc.} {\bf 485} (June, 2019) 5010--5020,
  [\href{http://www.arxiv.org/abs/1812.06026}{{\tt 1812.06026}}].

\bibitem{Cole:2000ex}
S.~Cole, C.~G. Lacey, C.~M. Baugh, and C.~S. Frenk, {\it {Hierarchical galaxy
  formation}},  {\em Mon. Not. Roy. Astron. Soc.} {\bf 319} (2000) 168,
  [\href{http://www.arxiv.org/abs/astro-ph/0007281}{{\tt astro-ph/0007281}}].

\bibitem{Parkinson:2007yh}
H.~Parkinson, S.~Cole, and J.~Helly, {\it {Generating Dark Matter Halo Merger
  Trees}},  {\em Mon. Not. Roy. Astron. Soc.} {\bf 383} (2008) 557,
  [\href{http://www.arxiv.org/abs/0708.1382}{{\tt 0708.1382}}].

\bibitem{Sheth:1999su}
R.~K. Sheth, H.~J. Mo, and G.~Tormen, {\it {Ellipsoidal collapse and an
  improved model for the number and spatial distribution of dark matter
  haloes}},  {\em Mon. Not. Roy. Astron. Soc.} {\bf 323} (2001) 1,
  [\href{http://www.arxiv.org/abs/astro-ph/9907024}{{\tt astro-ph/9907024}}].

\bibitem{2017MNRAS.471.1709G}
S.~{Garrison-Kimmel}, {\em et~al.}, {\it {Not so lumpy after all: modelling the
  depletion of dark matter subhaloes by Milky Way-like galaxies}},  {\em {Mon.
  Not. Roy. Astron. Soc.}} {\bf 471} (Oct., 2017) 1709--1727,
  [\href{http://www.arxiv.org/abs/1701.03792}{{\tt 1701.03792}}].

\bibitem{Benson:2019jio}
A.~J. Benson, {\it {The Normalization and Slope of the Dark Matter (Sub-)Halo
  Mass Function on Sub-Galactic Scales}},  {\em Mon. Not. Roy. Astron. Soc.}
  {\bf 493} (2020), no.~1 1268--1276,
  [\href{http://www.arxiv.org/abs/1911.04579}{{\tt 1911.04579}}].

\bibitem{2020MNRAS.492.5780R}
J.~{Richings}, {\em et~al.}, {\it {Subhalo destruction in the APOSTLE and
  AURIGA simulations}},  {\em {Mon. Not. Roy. Astron. Soc.}} {\bf 492} (Mar.,
  2020) 5780--5793, [\href{http://www.arxiv.org/abs/1811.12437}{{\tt
  1811.12437}}].

\bibitem{Samuel:2019ylk}
J.~Samuel {\em et~al.}, {\it {A profile in FIRE: resolving the radial
  distributions of satellite galaxies in the Local Group with simulations}},
  {\em Mon. Not. Roy. Astron. Soc.} {\bf 491} (2020), no.~1 1471--1490,
  [\href{http://www.arxiv.org/abs/1904.11508}{{\tt 1904.11508}}].

\bibitem{Navarro:1996gj}
J.~F. Navarro, C.~S. Frenk, and S.~D.~M. White, {\it {A Universal density
  profile from hierarchical clustering}},  {\em Astrophys. J.} {\bf 490} (1997)
  493--508, [\href{http://www.arxiv.org/abs/astro-ph/9611107}{{\tt
  astro-ph/9611107}}].

\bibitem{Wechsler:2001cs}
R.~H. Wechsler, J.~S. Bullock, J.~R. Primack, A.~V. Kravtsov, and A.~Dekel,
  {\it {Concentrations of dark halos from their assembly histories}},  {\em
  Astrophys. J.} {\bf 568} (2002) 52--70,
  [\href{http://www.arxiv.org/abs/astro-ph/0108151}{{\tt astro-ph/0108151}}].

\bibitem{Ludlow:2016ifl}
A.~D. Ludlow, S.~Bose, R.~E. Angulo, L.~Wang, W.~A. Hellwing, J.~F. Navarro,
  S.~Cole, and C.~S. Frenk, {\it {The
  mass\textendash{}concentration\textendash{}redshift relation of cold and warm
  dark matter haloes}},  {\em Mon. Not. Roy. Astron. Soc.} {\bf 460} (2016),
  no.~2 1214--1232, [\href{http://www.arxiv.org/abs/1601.02624}{{\tt
  1601.02624}}].

\bibitem{Diemer:2018vmz}
B.~Diemer and M.~Joyce, {\it {An accurate physical model for halo
  concentrations}},  {\em Astrophys. J.} {\bf 871} (2019), no.~2 168,
  [\href{http://www.arxiv.org/abs/1809.07326}{{\tt 1809.07326}}].

\bibitem{Wang:2019ftp}
J.~Wang, S.~Bose, C.~S. Frenk, L.~Gao, A.~Jenkins, V.~Springel, and S.~D.~M.
  White, {\it {Universal structure of dark matter haloes over a mass range of
  20 orders of magnitude}},  {\em Nature} {\bf 585} (2020), no.~7823 39--42,
  [\href{http://www.arxiv.org/abs/1911.09720}{{\tt 1911.09720}}].

\bibitem{Diemer:2014gba}
B.~Diemer and A.~V. Kravtsov, {\it {A universal model for halo
  concentrations}},  {\em Astrophys. J.} {\bf 799} (2015), no.~1 108,
  [\href{http://www.arxiv.org/abs/1407.4730}{{\tt 1407.4730}}].

\bibitem{Lacey:1993iv}
C.~G. Lacey and S.~Cole, {\it {Merger rates in hierarchical models of galaxy
  formation}},  {\em Mon. Not. Roy. Astron. Soc.} {\bf 262} (1993) 627--649.

\bibitem{Dodelson:2003ft}
S.~Dodelson, {\em {Modern Cosmology}}.
\newblock Academic Press, Amsterdam, 2003.

\bibitem{Deason:2015hla}
A.~J. Deason, A.~R. Wetzel, S.~Garrison-Kimmel, and V.~Belokurov, {\it
  {Satellites of LMC-mass dwarfs: close friendships ruined by Milky Way mass
  haloes}},  {\em Mon. Not. Roy. Astron. Soc.} {\bf 453} (2015), no.~4
  3568--3574, [\href{http://www.arxiv.org/abs/1504.04372}{{\tt 1504.04372}}].

\bibitem{Barry:2023ksd}
M.~Barry, A.~Wetzel, S.~Chapman, J.~Samuel, R.~Sanderson, and A.~Arora, {\it
  {The dark side of FIRE: predicting the population of dark matter subhaloes
  around Milky Way-mass galaxies}},  {\em Mon. Not. Roy. Astron. Soc.} {\bf
  523} (2023), no.~1 428--440, [\href{http://www.arxiv.org/abs/2303.05527}{{\tt
  2303.05527}}].

\bibitem{2021MNRAS.504.5270D}
R.~{D'Souza} and E.~F. {Bell}, {\it {The infall of dwarf satellite galaxies are
  influenced by their host's massive accretions}},  {\em Mon. Not. Roy. Astron.
  Soc.} {\bf 504} (July, 2021) 5270--5286,
  [\href{http://www.arxiv.org/abs/2104.13249}{{\tt 2104.13249}}].

\bibitem{DES:2019ltu}
{\bf DES} {\bf Collaboration}, E.~O. Nadler {\em et~al.}, {\it {Milky Way
  Satellite Census -- II. Galaxy-Halo Connection Constraints Including the
  Impact of the Large Magellanic Cloud}},  {\em Astrophys. J.} {\bf 893} (2020)
  48, [\href{http://www.arxiv.org/abs/1912.03303}{{\tt 1912.03303}}].

\bibitem{2021ApJ...920L..11N}
E.~O. {Nadler}, A.~{Banerjee}, S.~{Adhikari}, Y.-Y. {Mao}, and R.~H.
  {Wechsler}, {\it {The Effects of Dark Matter and Baryonic Physics on the
  Milky Way Subhalo Population in the Presence of the Large Magellanic Cloud}},
   {\em Astrophys. J. Lett.} {\bf 920} (Oct., 2021) L11,
  [\href{http://www.arxiv.org/abs/2109.12120}{{\tt 2109.12120}}].

\bibitem{2022MNRAS.516.3944M}
V.~{Manwadkar} and A.~V. {Kravtsov}, {\it {Forward-modelling the luminosity,
  distance, and size distributions of the Milky Way satellites}},  {\em Mon.
  Not. Roy. Astron. Soc.} {\bf 516} (Nov., 2022) 3944--3971,
  [\href{http://www.arxiv.org/abs/2112.04511}{{\tt 2112.04511}}].

\bibitem{2018ApJ...852...68C}
B.~C. {Conn}, H.~{Jerjen}, D.~{Kim}, and M.~{Schirmer}, {\it {On the Nature of
  Ultra-faint Dwarf Galaxy Candidates. I. DES1, Eridanus III, and Tucana V}},
  {\em Astrophys. J.} {\bf 852} (Jan., 2018) 68,
  [\href{http://www.arxiv.org/abs/1712.01439}{{\tt 1712.01439}}].

\bibitem{2018ApJ...857...70C}
B.~C. {Conn}, H.~{Jerjen}, D.~{Kim}, and M.~{Schirmer}, {\it {On the Nature of
  Ultra-faint Dwarf Galaxy Candidates. II. The Case of Cetus II}},  {\em
  Astrophys. J.} {\bf 857} (Apr., 2018) 70,
  [\href{http://www.arxiv.org/abs/1803.04563}{{\tt 1803.04563}}].

\bibitem{2018arXiv180902259J}
H.~{Jerjen}, B.~{Conn}, D.~{Kim}, and M.~{Schirmer}, {\it {On the Nature of
  Ultra-faint Dwarf Galaxy Candidates. III. Horologium I, Pictor I, Grus I, and
  Phoenix II}},  \href{http://www.arxiv.org/abs/1809.02259}{{\tt 1809.02259}}.

\bibitem{2018ApJ...863...25M}
B.~{Mutlu-Pakdil}, D.~J. {Sand}, J.~L. {Carlin}, K.~{Spekkens}, N.~{Caldwell},
  D.~{Crnojevi{\'c}}, A.~K. {Hughes}, B.~{Willman}, and D.~{Zaritsky}, {\it {A
  Deeper Look at the New Milky Way Satellites: Sagittarius II, Reticulum II,
  Phoenix II, and Tucana III}},  {\em \apj} {\bf 863} (Aug., 2018) 25,
  [\href{http://www.arxiv.org/abs/1804.08627}{{\tt 1804.08627}}].

\bibitem{DES:2020szo}
{\bf DES} {\bf Collaboration}, S.~A. Cantu {\em et~al.}, {\it {A Deeper Look at
  DES Dwarf Galaxy Candidates: Grus i and Indus ii}},  {\em Astrophys. J.} {\bf
  916} (2021), no.~2 81, [\href{http://www.arxiv.org/abs/2005.06478}{{\tt
  2005.06478}}].

\bibitem{DELVE:2019xvr}
{\bf DELVE} {\bf Collaboration}, S.~Mau {\em et~al.}, {\it {Two Ultra-Faint
  Milky Way Stellar Systems Discovered in Early Data from the DECam Local
  Volume Exploration Survey}},  {\em Astrophys. J.} {\bf 890} (2020), no.~2
  136, [\href{http://www.arxiv.org/abs/1912.03301}{{\tt 1912.03301}}].

\bibitem{DELVE:2021rcc}
{\bf DELVE} {\bf Collaboration}, C.~E. Mart\'\i{}nez-V\'azquez {\em et~al.},
  {\it {RR Lyrae Stars in the Newly Discovered Ultra-faint Dwarf Galaxy
  Centaurus I*}},  {\em Astron. J.} {\bf 162} (2021), no.~6 253,
  [\href{http://www.arxiv.org/abs/2107.05688}{{\tt 2107.05688}}].

\bibitem{DELVE:2021lds}
{\bf DELVE} {\bf Collaboration}, W.~Cerny {\em et~al.}, {\it {Eridanus IV: an
  Ultra-faint Dwarf Galaxy Candidate Discovered in the DECam Local Volume
  Exploration Survey}},  {\em Astrophys. J. Lett.} {\bf 920} (2021), no.~2 L44,
  [\href{http://www.arxiv.org/abs/2107.09080}{{\tt 2107.09080}}].

\bibitem{DELVE:2022ijm}
{\bf DELVE} {\bf Collaboration}, W.~Cerny {\em et~al.}, {\it {Pegasus IV:
  Discovery and Spectroscopic Confirmation of an Ultra-faint Dwarf Galaxy in
  the Constellation Pegasus}},  {\em Astrophys. J.} {\bf 942} (2023), no.~2
  111, [\href{http://www.arxiv.org/abs/2203.11788}{{\tt 2203.11788}}].

\bibitem{DELVE:2022jls}
{\bf DELVE} {\bf Collaboration}, W.~Cerny {\em et~al.}, {\it {Six More
  Ultra-Faint Milky Way Companions Discovered in the DECam Local Volume
  Exploration Survey}},  \href{http://www.arxiv.org/abs/2209.12422}{{\tt
  2209.12422}}.

\bibitem{Torrealba:2018svf}
G.~Torrealba {\em et~al.}, {\it {Discovery of two neighbouring satellites in
  the Carina constellation with MagLiteS}},  {\em Mon. Not. Roy. Astron. Soc.}
  {\bf 475} (2018), no.~4 5085--5097,
  [\href{http://www.arxiv.org/abs/1801.07279}{{\tt 1801.07279}}].

\bibitem{2016MNRAS.461.2212J}
P.~{Jethwa}, D.~{Erkal}, and V.~{Belokurov}, {\it {A Magellanic origin of the
  DES dwarfs}},  {\em Mon. Not. Roy. Astron. Soc.} {\bf 461} (Sept., 2016)
  2212--2233, [\href{http://www.arxiv.org/abs/1603.04420}{{\tt 1603.04420}}].

\bibitem{2021ApJ...923..140G}
N.~{Garavito-Camargo}, E.~{Patel}, G.~{Besla}, A.~M. {Price-Whelan}, F.~A.
  {G{\'o}mez}, C.~F.~P. {Laporte}, and K.~V. {Johnston}, {\it {The Clustering
  of Orbital Poles Induced by the LMC: Hints for the Origin of Planes of
  Satellites}},  {\em \apj} {\bf 923} (Dec., 2021) 140,
  [\href{http://www.arxiv.org/abs/2108.07321}{{\tt 2108.07321}}].

\bibitem{Vegetti:2014lqa}
S.~Vegetti, L.~V.~E. Koopmans, M.~W. Auger, T.~Treu, and A.~S. Bolton, {\it
  {Inference of the cold dark matter substructure mass function at z = 0.2
  using strong gravitational lenses}},  {\em Mon. Not. Roy. Astron. Soc.} {\bf
  442} (2014), no.~3 2017--2035,
  [\href{http://www.arxiv.org/abs/1405.3666}{{\tt 1405.3666}}].

\bibitem{Inoue:2014jka}
K.~T. Inoue, R.~Takahashi, T.~Takahashi, and T.~Ishiyama, {\it {Constraints on
  warm dark matter from weak lensing in anomalous quadruple lenses}},  {\em
  Mon. Not. Roy. Astron. Soc.} {\bf 448} (2015), no.~3 2704--2716,
  [\href{http://www.arxiv.org/abs/1409.1326}{{\tt 1409.1326}}].

\bibitem{Hezaveh:2016ltk}
Y.~D. Hezaveh {\em et~al.}, {\it {Detection of lensing substructure using ALMA
  observations of the dusty galaxy SDP.81}},  {\em Astrophys. J.} {\bf 823}
  (2016), no.~1 37, [\href{http://www.arxiv.org/abs/1601.01388}{{\tt
  1601.01388}}].

\bibitem{Birrer:2017rpp}
S.~Birrer, A.~Amara, and A.~Refregier, {\it {Lensing substructure
  quantification in RXJ1131-1231: A 2 keV lower bound on dark matter thermal
  relic mass}},  {\em JCAP} {\bf 05} (2017) 037,
  [\href{http://www.arxiv.org/abs/1702.00009}{{\tt 1702.00009}}].

\bibitem{Gilman:2019nap}
D.~Gilman, S.~Birrer, A.~Nierenberg, T.~Treu, X.~Du, and A.~Benson, {\it {Warm
  dark matter chills out: constraints on the halo mass function and the
  free-streaming length of dark matter with eight quadruple-image strong
  gravitational lenses}},  {\em Mon. Not. Roy. Astron. Soc.} {\bf 491} (2020),
  no.~4 6077--6101, [\href{http://www.arxiv.org/abs/1908.06983}{{\tt
  1908.06983}}].

\bibitem{Hsueh:2019ynk}
J.-W. Hsueh, W.~Enzi, S.~Vegetti, M.~Auger, C.~D. Fassnacht, G.~Despali,
  L.~V.~E. Koopmans, and J.~P. McKean, {\it {SHARP \textendash{} VII. New
  constraints on the dark matter free-streaming properties and substructure
  abundance from gravitationally lensed quasars}},  {\em Mon. Not. Roy. Astron.
  Soc.} {\bf 492} (2020), no.~2 3047--3059,
  [\href{http://www.arxiv.org/abs/1905.04182}{{\tt 1905.04182}}].

\bibitem{1993ApJ...403...74Q}
P.~J. {Quinn}, L.~{Hernquist}, and D.~P. {Fullagar}, {\it {Heating of Galactic
  Disks by Mergers}},  {\em Astrophys.~J.} {\bf 403} (Jan., 1993) 74.

\bibitem{Feldmann:2013hqa}
R.~Feldmann and D.~Spolyar, {\it {Detecting Dark Matter Substructures around
  the Milky Way with Gaia}},  {\em Mon. Not. Roy. Astron. Soc.} {\bf 446}
  (2015) 1000--1012, [\href{http://www.arxiv.org/abs/1310.2243}{{\tt
  1310.2243}}].

\bibitem{Yoon:2010iy}
J.~H. Yoon, K.~V. Johnston, and D.~W. Hogg, {\it {Clumpy Streams from Clumpy
  Halos: Detecting Missing Satellites with Cold Stellar Structures}},  {\em
  Astrophys. J.} {\bf 731} (2011) 58,
  [\href{http://www.arxiv.org/abs/1012.2884}{{\tt 1012.2884}}].

\bibitem{Carlberg:2011xj}
R.~G. Carlberg, {\it {Dark Matter Sub-Halo Counts via Star Stream Crossings}},
  {\em Astrophys. J.} {\bf 748} (2012) 20,
  [\href{http://www.arxiv.org/abs/1109.6022}{{\tt 1109.6022}}].

\bibitem{Aganze:2023nkp}
C.~Aganze, S.~Pearson, T.~Starkenburg, G.~Contardo, K.~V. Johnston,
  K.~Tavangar, A.~M. Price-Whelan, and A.~J. Burgasser, {\it {Prospects for
  Detecting Gaps in Globular Cluster Stellar Streams in External Galaxies with
  the Nancy Grace Roman Space Telescope}},
  \href{http://www.arxiv.org/abs/2305.12045}{{\tt 2305.12045}}.

\bibitem{Tollerud:2010bj}
E.~J. Tollerud, J.~S. Bullock, G.~J. Graves, and J.~Wolf, {\it {From Galaxy
  Clusters to Ultra-Faint Dwarf Spheroidals: A Fundamental Curve Connecting
  Dispersion-supported Galaxies to Their Dark Matter Halos}},  {\em Astrophys.
  J.} {\bf 726} (2011) 108, [\href{http://www.arxiv.org/abs/1007.5311}{{\tt
  1007.5311}}].

\bibitem{Strigari:2008ib}
L.~E. Strigari, J.~S. Bullock, M.~Kaplinghat, J.~D. Simon, M.~Geha, B.~Willman,
  and M.~G. Walker, {\it {A common mass scale for satellite galaxies of the
  Milky Way}},  {\em Nature} {\bf 454} (2008) 1096--1097,
  [\href{http://www.arxiv.org/abs/0808.3772}{{\tt 0808.3772}}].

\bibitem{McConnachie:2012vd}
A.~W. McConnachie, {\it {The observed properties of dwarf galaxies in and
  around the Local Group}},  {\em Astron. J.} {\bf 144} (2012) 4,
  [\href{http://www.arxiv.org/abs/1204.1562}{{\tt 1204.1562}}]. Updated as of
  2021 in
  \href{https://www.cadc-ccda.hia-iha.nrc-cnrc.gc.ca/en/community/nearby/}{this
  URL}.

\bibitem{Read:2017lvq}
J.~I. Read, G.~Iorio, O.~Agertz, and F.~Fraternali, {\it {The stellar
  mass\textendash{}halo mass relation of isolated field dwarfs: a critical test
  of \ensuremath{\Lambda}CDM at the edge of galaxy formation}},  {\em Mon. Not.
  Roy. Astron. Soc.} {\bf 467} (2017), no.~2 2019--2038,
  [\href{http://www.arxiv.org/abs/1607.03127}{{\tt 1607.03127}}].

\bibitem{Danieli:2017uvz}
S.~Danieli, P.~van Dokkum, and C.~Conroy, {\it {Hunting Faint Dwarf Galaxies in
  the Field Using Integrated Light Surveys}},  {\em Astrophys. J.} {\bf 856}
  (2018), no.~1 69, [\href{http://www.arxiv.org/abs/1711.00860}{{\tt
  1711.00860}}].

\bibitem{Jiang:2018ioo}
F.~Jiang {\em et~al.}, {\it {Is the dark-matter halo spin a predictor of galaxy
  spin and size?}},  {\em Mon. Not. Roy. Astron. Soc.} {\bf 488} (2019), no.~4
  4801--4815, [\href{http://www.arxiv.org/abs/1804.07306}{{\tt 1804.07306}}].

\bibitem{Moster:2012fv}
B.~P. Moster, T.~Naab, and S.~D.~M. White, {\it {Galactic star formation and
  accretion histories from matching galaxies to dark matter haloes}},  {\em
  Mon. Not. Roy. Astron. Soc.} {\bf 428} (2013) 3121,
  [\href{http://www.arxiv.org/abs/1205.5807}{{\tt 1205.5807}}].

\bibitem{Wechsler:2018pic}
R.~H. Wechsler and J.~L. Tinker, {\it {The Connection between Galaxies and
  their Dark Matter Halos}},  {\em Ann. Rev. Astron. Astrophys.} {\bf 56}
  (2018) 435--487, [\href{http://www.arxiv.org/abs/1804.03097}{{\tt
  1804.03097}}].

\bibitem{2021ApJ...923...35M}
F.~{Munshi}, A.~M. {Brooks}, E.~{Applebaum}, C.~R. {Christensen}, T.~{Quinn},
  and S.~{Sligh}, {\it {Quantifying Scatter in Galaxy Formation at the Lowest
  Masses}},  {\em \apj} {\bf 923} (Dec., 2021) 35,
  [\href{http://www.arxiv.org/abs/2101.05822}{{\tt 2101.05822}}].

\bibitem{2023MNRAS.519..871Z}
D.~{Zaritsky} and P.~{Behroozi}, {\it {Photometric mass estimation and the
  stellar mass-halo mass relation for low mass galaxies}},  {\em {Mon. Not.
  Roy. Astron. Soc.}} {\bf 519} (Feb., 2023) 871--883,
  [\href{http://www.arxiv.org/abs/2212.02948}{{\tt 2212.02948}}].

\bibitem{Garrison-Kimmel:2016szj}
S.~Garrison-Kimmel, J.~S. Bullock, M.~Boylan-Kolchin, and E.~Bardwell, {\it
  {Organized Chaos: Scatter in the relation between stellar mass and halo mass
  in small galaxies}},  {\em Mon. Not. Roy. Astron. Soc.} {\bf 464} (2017),
  no.~3 3108--3120, [\href{http://www.arxiv.org/abs/1603.04855}{{\tt
  1603.04855}}].

\bibitem{2019MNRAS.483.1314B}
T.~{Buck}, A.~V. {Macci{\`o}}, A.~A. {Dutton}, A.~{Obreja}, and J.~{Frings},
  {\it {NIHAO XV: the environmental impact of the host galaxy on galactic
  satellite and field dwarf galaxies}},  {\em {Mon. Not. Roy. Astron. Soc.}}
  {\bf 483} (Feb., 2019) 1314--1341,
  [\href{http://www.arxiv.org/abs/1804.04667}{{\tt 1804.04667}}].

\bibitem{Grand:2021fpx}
R.~J.~J. Grand, F.~Marinacci, R.~Pakmor, C.~M. Simpson, A.~J. Kelly, F.~A.
  G\'omez, A.~Jenkins, V.~Springel, C.~S. Frenk, and S.~D.~M. White, {\it
  {Determining the full satellite population of a Milky Way-mass halo in a
  highly resolved cosmological hydrodynamic simulation}},  {\em Mon. Not. Roy.
  Astron. Soc.} {\bf 507} (2021), no.~4 4953--4967,
  [\href{http://www.arxiv.org/abs/2105.04560}{{\tt 2105.04560}}].

\bibitem{Wolf:2009tu}
J.~Wolf, G.~D. Martinez, J.~S. Bullock, M.~Kaplinghat, M.~Geha, R.~R. Munoz,
  J.~D. Simon, and F.~F. Avedo, {\it {Accurate Masses for Dispersion-supported
  Galaxies}},  {\em Mon. Not. Roy. Astron. Soc.} {\bf 406} (2010) 1220,
  [\href{http://www.arxiv.org/abs/0908.2995}{{\tt 0908.2995}}].

\bibitem{2005MNRAS.356..107R}
J.~I. {Read} and G.~{Gilmore}, {\it {Mass loss from dwarf spheroidal galaxies:
  the origins of shallow dark matter cores and exponential surface brightness
  profiles}},  {\em Mon. Not. Roy. Astron. Soc.} {\bf 356} (Jan., 2005)
  107--124, [\href{http://www.arxiv.org/abs/astro-ph/0409565}{{\tt
  astro-ph/0409565}}].

\bibitem{Leaman:2012bi}
R.~Leaman, K.~A. Venn, A.~M. Brooks, G.~Battaglia, A.~A. Cole, R.~A. Ibata,
  M.~J. Irwin, A.~W. McConnachie, J.~T. Mendel, and E.~Tolstoy, {\it {The
  Resolved Structure and Dynamics of an Isolated Dwarf Galaxy: A VLT and Keck
  Spectroscopic Survey of WLM}},  {\em Astrophys. J.} {\bf 750} (2012) 33,
  [\href{http://www.arxiv.org/abs/1202.4474}{{\tt 1202.4474}}].

\bibitem{Weisz:2011mw}
D.~R. Weisz {\em et~al.}, {\it {Modeling the Effects of Star Formation
  Histories on Halpha and Ultra-Violet Fluxes in Nearby Dwarf Galaxies}},  {\em
  Astrophys. J.} {\bf 744} (2012) 44,
  [\href{http://www.arxiv.org/abs/1109.2905}{{\tt 1109.2905}}].

\bibitem{2012MNRAS.422.1231G}
F.~{Governato}, A.~{Zolotov}, A.~{Pontzen}, C.~{Christensen}, S.~H. {Oh}, A.~M.
  {Brooks}, T.~{Quinn}, S.~{Shen}, and J.~{Wadsley}, {\it {Cuspy no more: how
  outflows affect the central dark matter and baryon distribution in
  {\ensuremath{\Lambda}} cold dark matter galaxies}},  {\em {Mon. Not. Roy.
  Astron. Soc.}} {\bf 422} (May, 2012) 1231--1240,
  [\href{http://www.arxiv.org/abs/1202.0554}{{\tt 1202.0554}}].

\bibitem{Teyssier:2012ie}
R.~Teyssier, A.~Pontzen, Y.~Dubois, and J.~Read, {\it {Cusp-core
  transformations in dwarf galaxies: observational predictions}},  {\em Mon.
  Not. Roy. Astron. Soc.} {\bf 429} (2013) 3068,
  [\href{http://www.arxiv.org/abs/1206.4895}{{\tt 1206.4895}}].

\bibitem{2014MNRAS.437..415D}
A.~{Di Cintio}, C.~B. {Brook}, A.~V. {Macci{\`o}}, G.~S. {Stinson}, A.~{Knebe},
  A.~A. {Dutton}, and J.~{Wadsley}, {\it {The dependence of dark matter
  profiles on the stellar-to-halo mass ratio: a prediction for cusps versus
  cores}},  {\em {Mon. Not. Roy. Astron. Soc.}} {\bf 437} (Jan., 2014)
  415--423, [\href{http://www.arxiv.org/abs/1306.0898}{{\tt 1306.0898}}].

\bibitem{Kauffmann:2014cda}
G.~Kauffmann, {\it {Quantitative constraints on starburst cycles in galaxies
  with stellar masses in the range 10$^8$\textendash{}10$^{10}$ M $_\odot$}},
  {\em Mon. Not. Roy. Astron. Soc.} {\bf 441} (2014), no.~3 2717--2724,
  [\href{http://www.arxiv.org/abs/1401.8091}{{\tt 1401.8091}}].

\bibitem{McQuinn:2015pba}
K.~B.~W. McQuinn, F.~Lelli, E.~D. Skillman, A.~E. Dolphin, S.~S. McGaugh, and
  B.~F. Williams, {\it {The link between mass distribution and starbursts in
  dwarf galaxies}},  {\em Mon. Not. Roy. Astron. Soc.} {\bf 450} (2015), no.~4
  3886--3892, [\href{http://www.arxiv.org/abs/1504.02084}{{\tt 1504.02084}}].

\bibitem{Penarrubia:2012bb}
J.~Pe\~narrubia, A.~Pontzen, M.~G. Walker, and S.~E. Koposov, {\it {The
  coupling between the core/cusp and missing satellite problems}},  {\em
  Astrophys. J. Lett.} {\bf 759} (2012) L42,
  [\href{http://www.arxiv.org/abs/1207.2772}{{\tt 1207.2772}}].

\bibitem{Errani:2022aru}
R.~Errani, J.~F. Navarro, J.~Pe\~narrubia, B.~Famaey, and R.~Ibata, {\it {Dark
  matter halo cores and the tidal survival of Milky Way satellites}},  {\em
  Mon. Not. Roy. Astron. Soc.} {\bf 519} (2022), no.~1 384--396,
  [\href{http://www.arxiv.org/abs/2210.01131}{{\tt 2210.01131}}].

\bibitem{DiCintio:2013qxa}
A.~Di~Cintio, C.~B. Brook, A.~V. Macci\`o, G.~S. Stinson, A.~Knebe, A.~A.
  Dutton, and J.~Wadsley, {\it {The dependence of dark matter profiles on the
  stellar-to-halo mass ratio: a prediction for cusps versus cores}},  {\em Mon.
  Not. Roy. Astron. Soc.} {\bf 437} (2014), no.~1 415--423,
  [\href{http://www.arxiv.org/abs/1306.0898}{{\tt 1306.0898}}].

\bibitem{Bose:2018oaj}
S.~Bose {\em et~al.}, {\it {No cores in dark matter-dominated dwarf galaxies
  with bursty star formation histories}},  {\em Mon. Not. Roy. Astron. Soc.}
  {\bf 486} (2019), no.~4 4790--4804,
  [\href{http://www.arxiv.org/abs/1810.03635}{{\tt 1810.03635}}].

\bibitem{Read:2015sta}
J.~I. Read, O.~Agertz, and M.~L.~M. Collins, {\it {Dark matter cores all the
  way down}},  {\em Mon. Not. Roy. Astron. Soc.} {\bf 459} (2016), no.~3
  2573--2590, [\href{http://www.arxiv.org/abs/1508.04143}{{\tt 1508.04143}}].

\bibitem{Baltz:2007vq}
E.~A. Baltz, P.~Marshall, and M.~Oguri, {\it {Analytic models of plausible
  gravitational lens potentials}},  {\em JCAP} {\bf 01} (2009) 015,
  [\href{http://www.arxiv.org/abs/0705.0682}{{\tt 0705.0682}}].

\bibitem{King:1962wi}
I.~King, {\it {The structure of star clusters. I. An Empirical density law}},
  {\em Astron. J.} {\bf 67} (1962) 471.

\bibitem{Gnedin:1997vp}
O.~Y. Gnedin, L.~Hernquist, and J.~P. Ostriker, {\it {Tidal shocking by
  extended mass distributions}},  {\em Astrophys. J.} {\bf 514} (1999)
  109--118, [\href{http://www.arxiv.org/abs/astro-ph/9709161}{{\tt
  astro-ph/9709161}}].

\bibitem{Read:2005zm}
J.~I. Read, M.~I. Wilkinson, N.~W. Evans, G.~Gilmore, and J.~T. Kleyna, {\it
  {The Tidal stripping of satellites}},  {\em Mon. Not. Roy. Astron. Soc.} {\bf
  366} (2006) 429--437, [\href{http://www.arxiv.org/abs/astro-ph/0506687}{{\tt
  astro-ph/0506687}}].

\bibitem{Dooley:2016ajo}
G.~A. Dooley, A.~H.~G. Peter, M.~Vogelsberger, J.~Zavala, and A.~Frebel, {\it
  {Enhanced Tidal Stripping of Satellites in the Galactic Halo from Dark Matter
  Self-Interactions}},  {\em Mon. Not. Roy. Astron. Soc.} {\bf 461} (2016),
  no.~1 710--727, [\href{http://www.arxiv.org/abs/1603.08919}{{\tt
  1603.08919}}].

\bibitem{Delos:2019lik}
M.~S. Delos, {\it {Tidal evolution of dark matter annihilation rates in
  subhalos}},  {\em Phys. Rev. D} {\bf 100} (2019), no.~6 063505,
  [\href{http://www.arxiv.org/abs/1906.10690}{{\tt 1906.10690}}].

\bibitem{Drakos:2020ksc}
N.~E. Drakos, J.~E. Taylor, and A.~J. Benson, {\it {Mass loss in tidally
  stripped systems; the energy-based truncation method}},  {\em Mon. Not. Roy.
  Astron. Soc.} {\bf 494} (2020), no.~1 378--395,
  [\href{http://www.arxiv.org/abs/2003.09452}{{\tt 2003.09452}}].

\bibitem{Penarrubia:2010jk}
J.~Penarrubia, A.~J. Benson, M.~G. Walker, G.~Gilmore, A.~McConnachie, and
  L.~Mayer, {\it {The impact of dark matter cusps and cores on the satellite
  galaxy population around spiral galaxies}},  {\em Mon. Not. Roy. Astron.
  Soc.} {\bf 406} (2010) 1290, [\href{http://www.arxiv.org/abs/1002.3376}{{\tt
  1002.3376}}].

\bibitem{Errani:2020wgn}
R.~Errani and J.~F. Navarro, {\it {The asymptotic tidal remnants of cold dark
  matter subhaloes}},  {\em Mon. Not. Roy. Astron. Soc.} {\bf 505} (2021),
  no.~1 18--32, [\href{http://www.arxiv.org/abs/2011.07077}{{\tt 2011.07077}}].

\bibitem{2022MNRAS.517.1398B}
A.~J. {Benson} and X.~{Du}, {\it {Tidal tracks and artificial disruption of
  cold dark matter haloes}},  {\em {Mon. Not. Roy. Astron. Soc.}} {\bf 517}
  (Nov., 2022) 1398--1406, [\href{http://www.arxiv.org/abs/2206.01842}{{\tt
  2206.01842}}].

\bibitem{Garrison-Kimmel:2013eoa}
S.~Garrison-Kimmel, M.~Boylan-Kolchin, J.~Bullock, and K.~Lee, {\it {ELVIS:
  Exploring the Local Volume in Simulations}},  {\em Mon. Not. Roy. Astron.
  Soc.} {\bf 438} (2014), no.~3 2578--2596,
  [\href{http://www.arxiv.org/abs/1310.6746}{{\tt 1310.6746}}].

\bibitem{Simon:2019nxf}
J.~D. Simon, {\it {The Faintest Dwarf Galaxies}},  {\em Ann. Rev. Astron.
  Astrophys.} {\bf 57} (2019), no.~1 375--415,
  [\href{http://www.arxiv.org/abs/1901.05465}{{\tt 1901.05465}}].

\bibitem{2023ApJ...948...87W}
S.~{Weerasooriya}, M.~S. {Bovill}, A.~{Benson}, A.~M. {Musick}, and
  M.~{Ricotti}, {\it {Devouring the Milky Way Satellites: Modeling Dwarf
  Galaxies with Galacticus}},  {\em \apj} {\bf 948} (May, 2023) 87,
  [\href{http://www.arxiv.org/abs/2209.13663}{{\tt 2209.13663}}].

\bibitem{Santos-Santos:2021goz}
I.~M.~E. Santos-Santos, L.~V. Sales, A.~Fattahi, and J.~F. Navarro, {\it
  {Satellite mass functions and the faint end of the galaxy
  mass\textendash{}halo mass relation in LCDM}},  {\em Mon. Not. Roy. Astron.
  Soc.} {\bf 515} (2022), no.~3 3685--3697,
  [\href{http://www.arxiv.org/abs/2111.01158}{{\tt 2111.01158}}].

\bibitem{2018MNRAS.478.3879S}
J.~L. {Sanders}, N.~W. {Evans}, and W.~{Dehnen}, {\it {Tidal disruption of
  dwarf spheroidal galaxies: the strange case of Crater II}},  {\em Mon. Not.
  Roy. Astron. Soc.} {\bf 478} (Aug., 2018) 3879--3889,
  [\href{http://www.arxiv.org/abs/1802.09537}{{\tt 1802.09537}}].

\bibitem{Errani:2021rzi}
R.~Errani, J.~F. Navarro, R.~Ibata, and J.~Pe\~narrubia, {\it {Structure and
  kinematics of tidally limited satellite galaxies in LCDM}},  {\em Mon. Not.
  Roy. Astron. Soc.} {\bf 511} (2022), no.~4 6001--6018,
  [\href{http://www.arxiv.org/abs/2111.05866}{{\tt 2111.05866}}].

\bibitem{Dai:2019lud}
L.~Dai and J.~Miralda-Escud\'e, {\it {Gravitational Lensing Signatures of Axion
  Dark Matter Minihalos in Highly Magnified Stars}},  {\em Astron. J.} {\bf
  159} (2020), no.~2 49, [\href{http://www.arxiv.org/abs/1908.01773}{{\tt
  1908.01773}}].

\bibitem{2012ApJ...746..109L}
R.~{Lunnan}, M.~{Vogelsberger}, A.~{Frebel}, L.~{Hernquist}, A.~{Lidz}, and
  M.~{Boylan-Kolchin}, {\it {The Effects of Patchy Reionization on Satellite
  Galaxies of the Milky Way}},  {\em \apj} {\bf 746} (Feb., 2012) 109,
  [\href{http://www.arxiv.org/abs/1105.2293}{{\tt 1105.2293}}].

\bibitem{Koh:2016xmf}
D.~Koh and J.~H. Wise, {\it {Extending semi-numeric reionization models to the
  first stars and galaxies}},  {\em Mon. Not. Roy. Astron. Soc.} {\bf 474}
  (2018), no.~3 3817--3824, [\href{http://www.arxiv.org/abs/1609.04400}{{\tt
  1609.04400}}].

\bibitem{Rey:2019zsq}
M.~P. Rey, A.~Pontzen, O.~Agertz, M.~D.~A. Orkney, J.~I. Read, A.~Saintonge,
  and C.~Pedersen, {\it {EDGE: The origin of scatter in ultra-faint dwarf
  stellar masses and surface brightnesses}},  {\em Astrophys. J. Lett.} {\bf
  886} (2019), no.~1 L3, [\href{http://www.arxiv.org/abs/1909.04664}{{\tt
  1909.04664}}].

\bibitem{Barber:2013oua}
C.~Barber, E.~Starkenburg, J.~Navarro, A.~McConnachie, and A.~Fattahi, {\it
  {The Orbital Ellipticity of Satellite Galaxies and the Mass of the Milky
  Way}},  {\em Mon. Not. Roy. Astron. Soc.} {\bf 437} (2014), no.~1 959--967,
  [\href{http://www.arxiv.org/abs/1310.0466}{{\tt 1310.0466}}].

\bibitem{Woo:2008gg}
J.~Woo, S.~Courteau, and A.~Dekel, {\it {Scaling Relations and the Fundamental
  Line of the Local Group Dwarf Galaxies}},  {\em Mon. Not. Roy. Astron. Soc.}
  {\bf 390} (2008) 1453, [\href{http://www.arxiv.org/abs/0807.1331}{{\tt
  0807.1331}}].

\bibitem{Koposov:2007ni}
S.~Koposov {\em et~al.}, {\it {The Luminosity Function of the Milky Way
  Satellites}},  {\em Astrophys. J.} {\bf 686} (2008) 279--291,
  [\href{http://www.arxiv.org/abs/0706.2687}{{\tt 0706.2687}}].

\bibitem{Walsh:2008qn}
S.~Walsh, B.~Willman, and H.~Jerjen, {\it {The Invisibles: A Detection
  Algorithm to Trace the Faintest Milky Way Satellites}},  {\em Astron. J.}
  {\bf 137} (2009) 450, [\href{http://www.arxiv.org/abs/0807.3345}{{\tt
  0807.3345}}].

\bibitem{Garrison-Kimmel:2017zes}
S.~Garrison-Kimmel {\em et~al.}, {\it {Not so lumpy after all: modelling the
  depletion of dark matter subhaloes by Milky Way-like galaxies}},  {\em Mon.
  Not. Roy. Astron. Soc.} {\bf 471} (2017), no.~2 1709--1727,
  [\href{http://www.arxiv.org/abs/1701.03792}{{\tt 1701.03792}}].

\bibitem{Gnedin:2004cx}
O.~Y. Gnedin, A.~V. Kravtsov, A.~A. Klypin, and D.~Nagai, {\it {Response of
  dark matter halos to condensation of baryons: Cosmological simulations and
  improved adiabatic contraction model}},  {\em Astrophys. J.} {\bf 616} (2004)
  16--26, [\href{http://www.arxiv.org/abs/astro-ph/0406247}{{\tt
  astro-ph/0406247}}].

\bibitem{Abadi:2009ve}
M.~G. Abadi, J.~F. Navarro, M.~Fardal, A.~Babul, and M.~Steinmetz, {\it
  {Galaxy-Induced Transformation of Dark Matter Halos}},  {\em Mon. Not. Roy.
  Astron. Soc.} {\bf 407} (2010) 435--446,
  [\href{http://www.arxiv.org/abs/0902.2477}{{\tt 0902.2477}}].

\bibitem{DOnghia:2009xhq}
E.~D'Onghia, V.~Springel, L.~Hernquist, and D.~Keres, {\it {Substructure
  depletion in the Milky Way halo by the disk}},  {\em Astrophys. J.} {\bf 709}
  (2010) 1138--1147, [\href{http://www.arxiv.org/abs/0907.3482}{{\tt
  0907.3482}}].

\bibitem{Brooks:2012ah}
A.~M. Brooks, M.~Kuhlen, A.~Zolotov, and D.~Hooper, {\it {A Baryonic Solution
  to the Missing Satellites Problem}},  {\em Astrophys. J.} {\bf 765} (2013)
  22, [\href{http://www.arxiv.org/abs/1209.5394}{{\tt 1209.5394}}].

\bibitem{Sawala:2016tlo}
T.~Sawala, P.~Pihajoki, P.~H. Johansson, C.~S. Frenk, J.~F. Navarro, K.~A.
  Oman, and S.~D.~M. White, {\it {Shaken and Stirred: The Milky Way's Dark
  Substructures}},  {\em Mon. Not. Roy. Astron. Soc.} {\bf 467} (2017), no.~4
  4383--4400, [\href{http://www.arxiv.org/abs/1609.01718}{{\tt 1609.01718}}].

\bibitem{2022ApJ...940..136P}
A.~B. {Pace}, D.~{Erkal}, and T.~S. {Li}, {\it {Proper Motions, Orbits, and
  Tidal Influences of Milky Way Dwarf Spheroidal Galaxies}},  {\em Astrophys.
  J.} {\bf 940} (Dec., 2022) 136,
  [\href{http://www.arxiv.org/abs/2205.05699}{{\tt 2205.05699}}].

\bibitem{Kirby:2013isa}
E.~N. Kirby, M.~Boylan-Kolchin, J.~G. Cohen, M.~Geha, J.~S. Bullock, and
  M.~Kaplinghat, {\it {Segue 2: The Least Massive Galaxy}},  {\em Astrophys.
  J.} {\bf 770} (2013) 16, [\href{http://www.arxiv.org/abs/1304.6080}{{\tt
  1304.6080}}].

\bibitem{Planck:2018nkj}
{\bf Planck} {\bf Collaboration}, N.~Aghanim {\em et~al.}, {\it {Planck 2018
  results. I. Overview and the cosmological legacy of Planck}},  {\em Astron.
  Astrophys.} {\bf 641} (2020) A1,
  [\href{http://www.arxiv.org/abs/1807.06205}{{\tt 1807.06205}}].

\bibitem{DES:2017qwj}
{\bf DES} {\bf Collaboration}, M.~A. Troxel {\em et~al.}, {\it {Dark Energy
  Survey Year 1 results: Cosmological constraints from cosmic shear}},  {\em
  Phys. Rev. D} {\bf 98} (2018), no.~4 043528,
  [\href{http://www.arxiv.org/abs/1708.01538}{{\tt 1708.01538}}].

\bibitem{Chabanier:2019eai}
S.~Chabanier, M.~Millea, and N.~Palanque-Delabrouille, {\it {Matter power
  spectrum: from Ly$\alpha$ forest to CMB scales}},  {\em Mon. Not. Roy.
  Astron. Soc.} {\bf 489} (2019), no.~2 2247--2253,
  [\href{http://www.arxiv.org/abs/1905.08103}{{\tt 1905.08103}}].

\bibitem{StenDelos:2019xdk}
M.~Sten~Delos, T.~Linden, and A.~L. Erickcek, {\it {Breaking a dark degeneracy:
  The gamma-ray signature of early matter domination}},  {\em Phys. Rev. D}
  {\bf 100} (2019), no.~12 123546,
  [\href{http://www.arxiv.org/abs/1910.08553}{{\tt 1910.08553}}].

\bibitem{Delos:2023vfv}
M.~S. Delos, K.~Redmond, and A.~L. Erickcek, {\it {How an era of kination
  impacts substructure and the dark matter annihilation rate}},
  \href{http://www.arxiv.org/abs/2304.12336}{{\tt 2304.12336}}.

\bibitem{2022MNRAS.514.2667K}
A.~{Kravtsov} and V.~{Manwadkar}, {\it {GRUMPY: a simple framework for
  realistic forward modelling of dwarf galaxies}},  {\em {Mon. Not. Roy.
  Astron. Soc.}} {\bf 514} (Aug., 2022) 2667--2691,
  [\href{http://www.arxiv.org/abs/2106.09724}{{\tt 2106.09724}}].

\bibitem{Edge_DarkLight}
S.~Kim {\em et~al.}, {\it {EDGE: Predictable Scatter in the Stellar Mass--Halo
  Mass Relation of Dwarf Galaxies}}, . In preparation.

\bibitem{Kravtsov:2023oxa}
A.~Kravtsov and Z.~Wu, {\it {Densities and mass assembly histories of the Milky
  Way satellites are not a challenge to $\Lambda$CDM}},
  \href{http://www.arxiv.org/abs/2306.08674}{{\tt 2306.08674}}.

\bibitem{DES:2019vzn}
{\bf DES} {\bf Collaboration}, A.~Drlica-Wagner {\em et~al.}, {\it {Milky Way
  Satellite Census. I. The Observational Selection Function for Milky Way
  Satellites in DES Y3 and Pan-STARRS DR1}},  {\em Astrophys. J.} {\bf 893}
  (2020) 1, [\href{http://www.arxiv.org/abs/1912.03302}{{\tt 1912.03302}}].

\bibitem{LSSTScience:2009jmu}
{\bf LSST Science, LSST Project} {\bf Collaboration}, P.~A. Abell {\em et~al.},
  {\it {LSST Science Book, Version 2.0}},
  \href{http://www.arxiv.org/abs/0912.0201}{{\tt 0912.0201}}.

\bibitem{Mutlu-Pakdil:2021crk}
B.~Mutlu-Pakd\i{}l, D.~J. Sand, D.~Crnojevi\'c, A.~Drlica-Wagner, N.~Caldwell,
  P.~Guhathakurta, A.~C. Seth, J.~D. Simon, J.~Strader, and E.~Toloba, {\it
  {Resolved Dwarf Galaxy Searches within \textasciitilde{}5 Mpc with the Vera
  Rubin Observatory and Subaru Hyper Suprime-Cam}},  {\em Astrophys. J.} {\bf
  918} (2021), no.~2 [\href{http://www.arxiv.org/abs/2105.01658}{{\tt
  2105.01658}}].

\bibitem{Planck:2016mks}
{\bf Planck} {\bf Collaboration}, R.~Adam {\em et~al.}, {\it {Planck
  intermediate results. XLVII. Planck constraints on reionization history}},
  {\em Astron. Astrophys.} {\bf 596} (2016) A108,
  [\href{http://www.arxiv.org/abs/1605.03507}{{\tt 1605.03507}}].

\bibitem{Castellano:2022knz}
M.~Castellano, N.~Menci, and M.~Romanello, {\it {Constraints On Dark Matter
  From Reionization}},  {\em Frascati Phys. Ser.} {\bf 74} (2022) 209--224,
  [\href{http://www.arxiv.org/abs/2301.03854}{{\tt 2301.03854}}].

\bibitem{Furlanetto:2015apc}
S.~R. Furlanetto, {\it {The 21-cm Line as a Probe of Reionization}},
  \href{http://www.arxiv.org/abs/1511.01131}{{\tt 1511.01131}}.

\bibitem{2022ApJ...940L..14N}
R.~P. {Naidu}, {\em et~al.}, {\it {Two Remarkably Luminous Galaxy Candidates at
  z {\ensuremath{\approx}} 10-12 Revealed by JWST}},  {\em {Astrophys. J.
  Lett.}} {\bf 940} (Nov., 2022) L14,
  [\href{http://www.arxiv.org/abs/2207.09434}{{\tt 2207.09434}}].

\bibitem{2022ApJ...938L..15C}
M.~{Castellano}, {\em et~al.}, {\it {Early Results from GLASS-JWST. III. Galaxy
  Candidates at z 9-15}},  {\em {Astrophys. J. Lett.}} {\bf 938} (Oct., 2022)
  L15, [\href{http://www.arxiv.org/abs/2207.09436}{{\tt 2207.09436}}].

\bibitem{2023ApJ...948L..14C}
M.~{Castellano}, {\em et~al.}, {\it {Early Results from GLASS-JWST. XIX. A High
  Density of Bright Galaxies at z {\ensuremath{\approx}} 10 in the A2744
  Region}},  {\em Astrophys. J. Lett.} {\bf 948} (May, 2023) L14,
  [\href{http://www.arxiv.org/abs/2212.06666}{{\tt 2212.06666}}].

\bibitem{2023MNRAS.518.4755A}
N.~J. {Adams}, {\em et~al.}, {\it {Discovery and properties of ultra-high
  redshift galaxies (9 < z < 12) in the JWST ERO SMACS 0723 Field}},  {\em
  {Mon. Not. Roy. Astron. Soc.}} {\bf 518} (Jan., 2023) 4755--4766,
  [\href{http://www.arxiv.org/abs/2207.11217}{{\tt 2207.11217}}].

\bibitem{2023MNRAS.519.1201A}
H.~{Atek}, {\em et~al.}, {\it {Revealing galaxy candidates out to z 16 with
  JWST observations of the lensing cluster SMACS0723}},  {\em {Mon. Not. Roy.
  Astron. Soc.}} {\bf 519} (Feb., 2023) 1201--1220,
  [\href{http://www.arxiv.org/abs/2207.12338}{{\tt 2207.12338}}].

\bibitem{2023MNRAS.519.3691D}
F.~R. {Donnan}, I.~{Garc{\'\i}a-Bernete}, D.~{Rigopoulou},
  M.~{Pereira-Santaella}, A.~{Alonso-Herrero}, P.~F. {Roche},
  A.~{Hern{\'a}n-Caballero}, and H.~W.~W. {Spoon}, {\it {The obscured nucleus
  and shocked environment of VV 114E revealed by JWST/MIRI spectroscopy}},
  {\em {Mon. Not. Roy. Astron. Soc.}} {\bf 519} (Mar., 2023) 3691--3705,
  [\href{http://www.arxiv.org/abs/2210.04647}{{\tt 2210.04647}}].

\bibitem{2023MNRAS.520.4554D}
C.~T. {Donnan}, D.~J. {McLeod}, R.~J. {McLure}, J.~S. {Dunlop}, A.~C.
  {Carnall}, F.~{Cullen}, and D.~{Magee}, {\it {The abundance of z
  {\ensuremath{\gtrsim}} 10 galaxy candidates in the HUDF using deep JWST
  NIRCam medium-band imaging}},  {\em {Mon. Not. Roy. Astron. Soc.}} {\bf 520}
  (Apr., 2023) 4554--4561, [\href{http://www.arxiv.org/abs/2212.10126}{{\tt
  2212.10126}}].

\bibitem{2023ApJS..265....5H}
Y.~{Harikane}, M.~{Ouchi}, M.~{Oguri}, Y.~{Ono}, K.~{Nakajima}, Y.~{Isobe},
  H.~{Umeda}, K.~{Mawatari}, and Y.~{Zhang}, {\it {A Comprehensive Study of
  Galaxies at z 9-16 Found in the Early JWST Data: Ultraviolet Luminosity
  Functions and Cosmic Star Formation History at the Pre-reionization Epoch}},
  {\em Astrophys. J. Suppl.} {\bf 265} (Mar., 2023) 5,
  [\href{http://www.arxiv.org/abs/2208.01612}{{\tt 2208.01612}}].

\bibitem{2023MNRAS.523.1009B}
R.~{Bouwens}, G.~{Illingworth}, P.~{Oesch}, M.~{Stefanon}, R.~{Naidu}, I.~{van
  Leeuwen}, and D.~{Magee}, {\it {UV luminosity density results at z > 8 from
  the first JWST/NIRCam fields: limitations of early data sets and the need for
  spectroscopy}},  {\em {Mon. Not. Roy. Astron. Soc.}} {\bf 523} (July, 2023)
  1009--1035, [\href{http://www.arxiv.org/abs/2212.06683}{{\tt 2212.06683}}].

\bibitem{2023MNRAS.523.1036B}
R.~J. {Bouwens}, M.~{Stefanon}, G.~{Brammer}, P.~A. {Oesch},
  T.~{Herard-Demanche}, G.~D. {Illingworth}, J.~{Matthee}, R.~P. {Naidu}, P.~G.
  {van Dokkum}, and I.~F. {van Leeuwen}, {\it {Evolution of the UV LF from z 15
  to z 8 using new JWST NIRCam medium-band observations over the HUDF/XDF}},
  {\em {Mon. Not. Roy. Astron. Soc.}} {\bf 523} (July, 2023) 1036--1055,
  [\href{http://www.arxiv.org/abs/2211.02607}{{\tt 2211.02607}}].

\bibitem{2022ApJ...938..144M}
T.~{Morishita}, {Abdurro'uf}, H.~{Hirashita}, A.~B. {Newman}, M.~{Stiavelli},
  and M.~{Chiaberge}, {\it {Compact Dust Emission in a Gravitationally Lensed
  Massive Quiescent Galaxy at z = 2.15 Revealed in 130 pc Resolution
  Observations by the Atacama Large Millimeter/submillimeter Array}},  {\em
  \apj} {\bf 938} (Oct., 2022) 144,
  [\href{http://www.arxiv.org/abs/2208.10525}{{\tt 2208.10525}}].

\bibitem{2023arXiv230406658H}
Y.~{Harikane}, K.~{Nakajima}, M.~{Ouchi}, H.~{Umeda}, Y.~{Isobe}, Y.~{Ono},
  Y.~{Xu}, and Y.~{Zhang}, {\it {Pure Spectroscopic Constraints on UV
  Luminosity Functions and Cosmic Star Formation History From 25 Galaxies at
  $z_\mathrm{spec}=8.61-13.20$ Confirmed with JWST/NIRSpec}},  {\em arXiv
  e-prints} (Apr., 2023) arXiv:2304.06658,
  [\href{http://www.arxiv.org/abs/2304.06658}{{\tt 2304.06658}}].

\bibitem{Dayal:2023nwi}
P.~Dayal and S.~K. Giri, {\it {Warm dark matter constraints from the JWST}},
  \href{http://www.arxiv.org/abs/2303.14239}{{\tt 2303.14239}}.

\bibitem{Menci:2016eui}
N.~Menci, A.~Grazian, M.~Castellano, and N.~G. Sanchez, {\it {A Stringent Limit
  on the Warm Dark Matter Particle Masses from the Abundance of z=6 Galaxies in
  the Hubble Frontier Fields}},  {\em Astrophys. J. Lett.} {\bf 825} (2016),
  no.~1 L1, [\href{http://www.arxiv.org/abs/1606.02530}{{\tt 1606.02530}}].

\bibitem{Sabti:2023xwo}
N.~Sabti, J.~B. Mu\~noz, and M.~Kamionkowski, {\it {Insights from HST into
  Ultra-Massive Galaxies and Early-Universe Cosmology}},
  \href{http://www.arxiv.org/abs/2305.07049}{{\tt 2305.07049}}.

\bibitem{Parashari:2023cui}
P.~Parashari and R.~Laha, {\it {Primordial power spectrum in light of JWST
  observations of high redshift galaxies}},
  \href{http://www.arxiv.org/abs/2305.00999}{{\tt 2305.00999}}.

\bibitem{Zwicky:1933gu}
F.~Zwicky, {\it {Die Rotverschiebung von extragalaktischen Nebeln}},  {\em
  Helv. Phys. Acta} {\bf 6} (1933) 110--127.

\bibitem{Zwicky:1937zza}
F.~Zwicky, {\it {On the Masses of Nebulae and of Clusters of Nebulae}},  {\em
  Astrophys. J.} {\bf 86} (1937) 217--246.

\bibitem{WMAP:2008lyn}
{\bf WMAP} {\bf Collaboration}, E.~Komatsu {\em et~al.}, {\it {Five-Year
  Wilkinson Microwave Anisotropy Probe (WMAP) Observations: Cosmological
  Interpretation}},  {\em Astrophys. J. Suppl.} {\bf 180} (2009) 330--376,
  [\href{http://www.arxiv.org/abs/0803.0547}{{\tt 0803.0547}}].

\bibitem{Zentner:2004dq}
A.~R. Zentner, A.~A. Berlind, J.~S. Bullock, A.~V. Kravtsov, and R.~H.
  Wechsler, {\it {The Physics of galaxy clustering. 1. A Model for subhalo
  populations}},  {\em Astrophys. J.} {\bf 624} (2005) 505--525,
  [\href{http://www.arxiv.org/abs/astro-ph/0411586}{{\tt astro-ph/0411586}}].

\bibitem{2018MNRAS.481.5073E}
R.~{Errani}, J.~{Pe{\~n}arrubia}, and M.~G. {Walker}, {\it {Systematics in
  virial mass estimators for pressure-supported systems}},  {\em Mon. Not. Roy.
  Astron. Soc.} {\bf 481} (Dec., 2018) 5073--5090,
  [\href{http://www.arxiv.org/abs/1805.00484}{{\tt 1805.00484}}].

\bibitem{Einasto:1965czb}
J.~Einasto, {\it {On the Construction of a Composite Model for the Galaxy and
  on the Determination of the System of Galactic Parameters}},  {\em Trudy
  Astrofizicheskogo Instituta Alma-Ata} {\bf 5} (1965) 87--100.

\bibitem{Nadler:2018iux}
E.~O. Nadler, Y.-Y. Mao, G.~M. Green, and R.~H. Wechsler, {\it {Modeling the
  Connection between Subhalos and Satellites in Milky Way\textendash{}like
  Systems}},  {\em Astrophys. J.} {\bf 873} (2019), no.~1 34,
  [\href{http://www.arxiv.org/abs/1809.05542}{{\tt 1809.05542}}].

\bibitem{Boylan-Kolchin:2009ztl}
M.~Boylan-Kolchin, V.~Springel, S.~D.~M. White, and A.~Jenkins, {\it {There's
  no place like home? Statistics of Milky Way-mass dark matter halos}},  {\em
  Mon. Not. Roy. Astron. Soc.} {\bf 406} (2010) 896,
  [\href{http://www.arxiv.org/abs/0911.4484}{{\tt 0911.4484}}].

\bibitem{Fielder:2018szt}
C.~E. Fielder, Y.-Y. Mao, J.~A. Newman, A.~R. Zentner, and T.~C. Licquia, {\it
  {Predictably missing satellites: subhalo abundances in Milky Way-like
  haloes}},  {\em Mon. Not. Roy. Astron. Soc.} {\bf 486} (2019), no.~4
  4545--4568, [\href{http://www.arxiv.org/abs/1807.05180}{{\tt 1807.05180}}].

\bibitem{Wilks:1938dza}
S.~S. Wilks, {\it {The Large-Sample Distribution of the Likelihood Ratio for
  Testing Composite Hypotheses}},  {\em Annals Math. Statist.} {\bf 9} (1938),
  no.~1 60--62.

\bibitem{Lepage:2020tgj}
G.~P. Lepage, {\it {Adaptive multidimensional integration: VEGAS enhanced}},
  {\em J. Comput. Phys.} {\bf 439} (2021) 110386,
  [\href{http://www.arxiv.org/abs/2009.05112}{{\tt 2009.05112}}].

\bibitem{bobyqa}
M.~J.~D. Powell, {\it {The BOBYQA algorithm for bound constrained optimization
  without derivatives}},  Tech. Rep. DAMTP 2009/NA06, University of Cambridge,
  2009.

\bibitem{10.1145/3338517}
C.~Cartis, J.~Fiala, B.~Marteau, and L.~Roberts, {\it Improving the flexibility
  and robustness of model-based derivative-free optimization solvers},  {\em
  ACM Trans. Math. Softw.} {\bf 45} (aug, 2019).

\bibitem{Belokurov:2009hw}
V.~Belokurov, M.~G. Walker, N.~W. Evans, G.~Gilmore, M.~J. Irwin, M.~Mateo,
  L.~Mayer, E.~Olszewski, J.~Bechtold, and T.~Pickering, {\it {Segue 2: A
  Prototype of the Population of Satellites of Satellites}},  {\em Mon. Not.
  Roy. Astron. Soc.} {\bf 397} (2009) 1748--1755,
  [\href{http://www.arxiv.org/abs/0903.0818}{{\tt 0903.0818}}].

\end{thebibliography}\endgroup

\clearpage
\onecolumngrid
\appendix
\renewcommand\thefigure{\thesection\arabic{figure}}

\setcounter{figure}{0}
\section{Impact of tidal stripping, subhalo infall times, the velocity estimator, and baryonic feedback}
\label{sec:app_baryon}
In this Appendix, we quantify the approximations described in the main text. We show that including tidal stripping, the different subhalo infall times, a different $\sigma_\mathrm{los}^*$ estimator, or a different DM halo profile affects our observables less than the observational uncertainties and than the parameters that we set free ---baryonic feedback that turns cusps into cores, the stellar mass-halo mass relation, the halo occupation fraction, and subhalo tidal disruption. If constraints on the latter are improved in future work and they stop dominating the error budget, these approximations may need revisiting.

\subsection{Tidal stripping}

In the main text, we neglect tidal stripping of subhalos by the Milky Way.

\begin{figure}[hbtp]
    \centering
    \includegraphics[width=0.49\textwidth, valign=t]{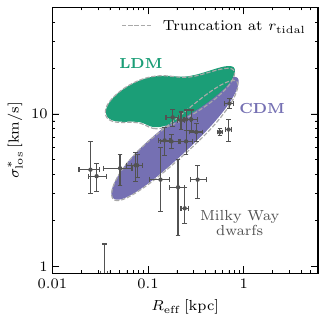}
    \includegraphics[width=0.49\textwidth, valign=t]{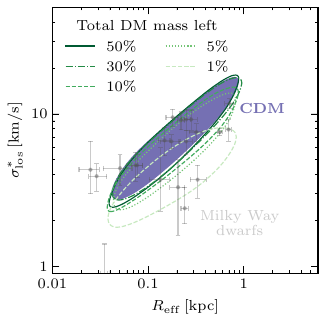}

    \caption{Same as \cref{fig:data}, but the dashed lines include tidal effects. To avoid crowding the figure, colored regions show 68\% of the enclosed population. In the left panel, we truncate DM density profiles at the tidal radius. In the right panel, we include tidal shocking and heating by changing the halo structural parameters, following the ``tidal tracks'' in Ref.~\cite{Penarrubia:2010jk}. \emph{Tidal effects would have a subleading impact on our error budget}.}
    \label{fig:tidal}
\end{figure}

\Cref{fig:tidal} shows that tidal effects have a subleading impact on our observables. Both panels show the correlation between $\sigma_\mathrm{los}^*$ and $R_\mathrm{eff}$, together with data from classical and SDSS satellites, for the same halo-galaxy connection model and LDM parameters as \cref{fig:data}, but including tidal effects in the dashed regions. In the left panel, we truncate the halo density profile beyond the tidal radius~\cite{Baltz:2007vq, Gilman:2019nap, Gilman:2021gkj}, as computed with \texttt{Galacticus} following Ref.~\cite{Zentner:2004dq} (we use the orbiting subhalo model, that follows subhalo positions and allows computing the tidal radius). In the right panel, we include tidal shocking and heating effects by changing halo structural parameters following the ``tidal tracks'' in Ref.~\cite{Penarrubia:2010jk}. We show only the CDM scenario to which Ref.~\cite{Penarrubia:2010jk} was calibrated. As mentioned in the main text, high concentrations of LDM halos may make them more resilient to tidal stripping~\cite{Dooley:2016ajo, Dai:2019lud}, so similar or smaller effects are expected for LDM.

We observe that DM mass removal beyond the tidal radius has a negligible impact on our observables given the precision of the data (see \cref{fig:extreme} below for the impact of the parameters that we set free). Tidal shocking and heating could have a significant impact, but this would require a large fraction of the DM halo population to lose more than $95\%$--$99\%$ of their mass. Simulations~\cite{Garrison-Kimmel:2013eoa, 2023ApJ...948...87W} and observations~\cite{Simon:2019nxf} disfavor such dramatic effects. Similar conclusions were reached in previous population studies: dwarf galaxies are located well inside their host DM halos, so very strong tidal stripping at the population level is needed to affect the conclusions~\cite{Kim:2021zzw, 2018MNRAS.478.3879S, Errani:2021rzi}.

\subsection{Infall times}

In the main text, we evaluate all Milky Way satellite properties at median infall redshift $z_\mathrm{infall} = 1$~\cite{Rocha:2011aa, Dooley:2016xkj, 2019arXiv190604180F}. 

\begin{figure}[hbtp]
    \centering
    \includegraphics[width=0.49\textwidth, valign=t]{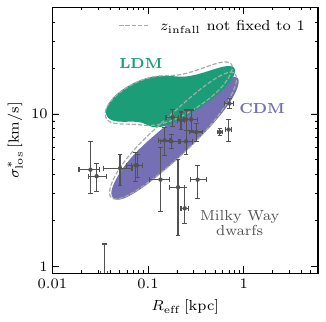}
    \caption{Same as \cref{fig:data}, but the dashed lines include the different satellite infall times. To avoid crowding the figure, colored regions show 68\% of the enclosed population. \emph{Different infall times would have a subleading impact on our error budget}.}
    \label{fig:zin}
\end{figure}

\Cref{fig:zin} shows that including the different infall redshifts has a subleading impact on our observables. We show the correlation between $\sigma_\mathrm{los}^*$ and $R_\mathrm{eff}$, together with data from classical and SDSS satellites, for the same halo-galaxy connection model and LDM parameters as \cref{fig:data}, but evaluating halo properties at infall redshift as computed with \texttt{Galacticus} in dashed. 

We observe that the impact on our observables is negligible  given the precision of the data (see \cref{fig:extreme} below for the impact of the parameters that we set free). The main effect is to slightly increase $\sigma_\mathrm{los}^*$. This is because the $z_\mathrm{infall}$ distribution is skewed towards $z_\mathrm{infall} > 1$~\cite{Rocha:2011aa, Dooley:2016xkj, 2019arXiv190604180F}, and internal halo velocities are higher at high redshift~\cite{Kim:2021zzw}.

\subsection{Velocity estimator}

In the main text, we estimate $\sigma_\mathrm{los}^*$ using \cref{eq:sigma_los}, that was obtained in Ref.~\cite{Wolf:2009tu}. This was shown to be robust against stellar velocity dispersion anisotropy. Ref.~\cite{2018MNRAS.481.5073E} proposed a different estimator, also robust against the shape of the inner halo profile and how deeply the stellar component is embedded within the halo. By comparing with simulation, the latter estimator was shown to have an accuracy of $\sim 10\%$ (better than our observational uncertainties).

\begin{figure}[hbtp]
    \centering
    \includegraphics[width=0.46\textwidth, valign=t]{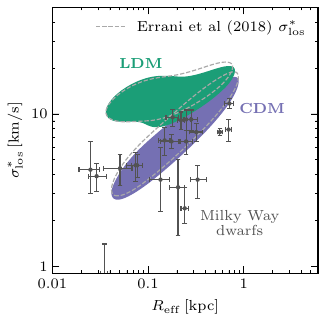}
    \caption{Same as \cref{fig:data}, but the dashed lines use the $\sigma_\mathrm{los}^*$ estimator from Ref.~\cite{2018MNRAS.481.5073E}. To avoid crowding the figure, colored regions show 68\% of the enclosed population. \emph{A recent mass estimator would have a subleading impact on our error budget}.}
    \label{fig:sigma_estimator}
\end{figure}

\Cref{fig:sigma_estimator} shows that using the estimator from Ref.~\cite{2018MNRAS.481.5073E} has a subleading impact on our observables. We show the correlation between $\sigma_\mathrm{los}^*$ and $R_\mathrm{eff}$, together with data from classical and SDSS satellites, for the same halo-galaxy connection model and LDM parameters as \cref{fig:data}, but with the estimator from Ref.~\cite{2018MNRAS.481.5073E} in the dashed regions. 

We observe that the impact on our observables is negligible  given the precision of the data (see \cref{fig:extreme} below for the impact of the parameters that we set free).

\subsection{DM halo profile}

In the main text, we assume that, before baryonic feedback induces cores, DM halo profiles follow an NFW form, \cref{eq:NFW}.

\begin{figure}[hbtp]
    \centering
    \includegraphics[width=0.49\textwidth, valign=t]{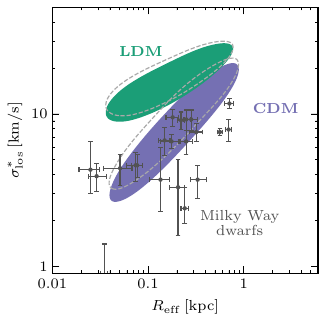}
    \caption{Same as \cref{fig:data}, but the dashed lines assume an Einasto profile and the solid region an NFW profile. To avoid crowding the figure, colored regions show 68\% of the enclosed population. \emph{Assuming an Einasto profile would have a subleading impact on our error budget}.}
    \label{fig:einasto}
\end{figure}

\Cref{fig:einasto} shows that assuming an Einasto form~\cite{Einasto:1965czb}
\begin{equation}
    \rho(r) = \rho_s e^{-\frac{2}{\alpha}\left[(r/r_s)^\alpha - 1\right]}
\end{equation}
has a subleading impact on our observables. We show the correlation between $\sigma_\mathrm{los}^*$ and $R_\mathrm{eff}$, together with data from classical and SDSS satellites. We use the same halo-galaxy connection model and LDM parameters as \cref{fig:data}, but assuming an Einasto profile in dashed and, for consistency, a cuspy NFW profile at all halo masses in solid. For the Einasto profile, we set $\alpha=0.16$, which was found in Ref.~\cite{Wang:2019ftp} to be a good fit for a very wide range in halo masses.

We observe that the impact on our observables is negligible  given the precision of the data (see \cref{fig:extreme} below for the impact of the parameters that we set free). The main effect is to slightly increase $\sigma_\mathrm{los}^*$. This is because for radii $\gtrsim 10^{-3} r_s$ an Einasto profile encloses more mass than an NFW profile with the same mass and concentration, and the half-light radii we consider are larger than $10^{-3} r_s$.

\subsection{Baryonic feedback}

Here, we illustrate the effects of some parameters that we set free in our analysis: baryonic feedback that turns NFW ``cusps'' into ``cores'', the uncertain stellar mass-halo mass relation, and the uncertain halo occupation fraction. As discussed in the main text, the first effect is the most degenerate with our determination of the power spectrum.

\begin{figure}[hbtp]
    \centering
    \includegraphics[width=0.46\textwidth, valign=t]{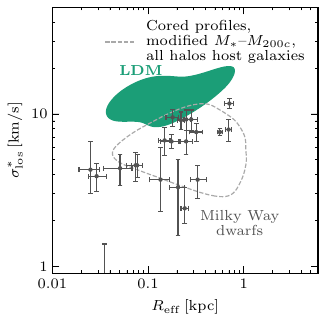}
    \caption{Same as \cref{fig:data}, but for the dashed line we change our free parameters. To avoid crowding the figure, colored regions show 68\% of the enclosed population for the LDM case. \emph{Our error budget is dominated by the parameters we set free}.}
    \label{fig:extreme}
\end{figure}

\Cref{fig:extreme} shows that the parameters that we set free in our analysis significantly affect our observables. We show the correlation between $\sigma_\mathrm{los}^*$ and $R_\mathrm{eff}$, together with data from classical and SDSS satellites, for the same halo-galaxy connection model and LDM parameters as \cref{fig:data}, but changing our free parameters in the dashed region. To generate the dashed region we set cored profiles for all halo masses, we change the scatter of the stellar mass-halo mass relation by setting $\sigma^{M_*}=0.8$ and $\gamma^{M_*}=-0.6$, and we set the halo occupation fraction to 1 at all halo masses. Importantly, these parameters are allowed by our analysis. Other parameters are set as in \cref{fig:data}.

We observe that, even if changing our free parameters has a strong impact, the LDM correlation between $\sigma_\mathrm{los}^*$ and $R_\mathrm{eff}$ is still different from CDM. For this correlation to be affected by baryons, \emph{baryonic feedback would have to induce increasingly strong cores at lower halo masses}, compensating the enhanced LDM mass-concentration relation in \cref{fig:physics}. This would be opposite to conventional wisdom, simulation, and observations --- lower mass halos tend to be increasingly ``cuspy'' as they have too few stars to instigate core formation~\cite{Penarrubia:2012bb, Errani:2022aru, Read:2015sta}.
\clearpage
\setcounter{figure}{0}
\section{Statistical analysis}
\label{sec:app_statistics}

We obtain the quantitative conclusions in the main text through a statistical analysis of Milky Way satellite properties. In this Appendix, we describe in detail this analysis.

We fit for the half-light radius, stellar velocity dispersion, and total number of classical and SDSS Milky Way satellites, using the data collected in Refs.~\cite{McConnachie:2012vd, Kim:2021zzw, Kirby:2013isa} and updated as of 2021 in \href{https://www.cadc-ccda.hia-iha.nrc-cnrc.gc.ca/en/community/nearby/}{this URL}. For reference, we compile the numerical values in \cref{sec:app_data}. We do not include the Magellanic clouds, Pisces II, and Sagittarius, as baryonic effects and strong tidal stripping could bias our results (the impact of this is minor, as they are a small fraction of our sample). We carry out a frequentist unbinned maximum likelihood (also called profile likelihood in the literature),
\begin{equation}
    -2 \ln \mathcal{L} = -2 \sum_i \ln \mu(\sigma_\mathrm{los}^{*\,\mathrm{obs,}\,i}, R_\mathrm{eff}^{\mathrm{obs,}\,i}) - 2 \ln \mathcal{P}(N_\mathrm{obs}) \, . \label{eq:likelihood}
\end{equation}
$\mu(\sigma_\mathrm{los}^{*\,\mathrm{obs}}, R_\mathrm{eff}^\mathrm{obs})\,\mathrm{d}\sigma_\mathrm{los}^{*\,\mathrm{obs}}\mathrm{d}R_\mathrm{eff}^\mathrm{obs}$ is the differential probability for a satellite galaxy to have an \emph{observed} stellar velocity dispersion $\sigma_\mathrm{los}^{*\,\mathrm{obs}}$ and half-light radius $R_\mathrm{eff}^\mathrm{obs}$. $\sigma_\mathrm{los}^{*\,\mathrm{obs,}\,i}$ and $R_\mathrm{eff}^{\mathrm{obs,}\,i}$ are the observed stellar velocity dispersions and half-light radii, respectively. $\mathcal{P}(N_\mathrm{obs})$ is the probability to observe $N_\mathrm{obs}$ satellite galaxies, with $N_\mathrm{obs} = 21$ the number of observed satellites. The first probability is given by
\begin{equation}
\label{eq:prob_final}
    \mu(\sigma_\mathrm{los}^{*\,\mathrm{obs}}, R_\mathrm{eff}^\mathrm{obs}) = \int \mathcal{P}(\sigma_\mathrm{los}^*, R_\mathrm{eff}, c, M_*, M) \frac{1}{\sqrt{2\pi} \Delta \sigma} e^{-\frac{(\sigma_\mathrm{los}^{*\,\mathrm{obs}} - \sigma_\mathrm{los}^*)^2}{2 \Delta \sigma^2}} \, \frac{1}{\sqrt{2\pi} \Delta R_\mathrm{eff}} e^{-\frac{(R_\mathrm{eff}^\mathrm{obs} - R_\mathrm{eff})^2}{2 \Delta R_\mathrm{eff}^2}} \mathrm{d}\sigma_\mathrm{los}^* \, \mathrm{d}R_\mathrm{eff} \, \mathrm{d}c\,\mathrm{d}M_* \,\mathrm{d}M \, .
\end{equation}
$\mathcal{P}(\sigma_\mathrm{los}^*, R_\mathrm{eff}, c, M_*, M)\,\mathrm{d}\sigma_\mathrm{los}^* \, \mathrm{d}R_\mathrm{eff} \, \mathrm{d}c\,\mathrm{d}M_* \,\mathrm{d}M$ is the differential probability to observe a galaxy with \emph{true} velocity dispersion $\sigma_\mathrm{los}^*$, half-light radius $R_\mathrm{eff}$, concentration $c$, stellar mass $M_*$, and DM halo mass $M$ (for simplicity, in this Appendix we denote $M_{200c} \equiv M$). The two Gaussians are the probabilities to observe a velocity dispersion $\sigma_\mathrm{los}^{*\,\mathrm{obs}}$ and half-light radius $R_\mathrm{eff}^\mathrm{obs}$, given their true values $\sigma_\mathrm{los}^*$ and $R_\mathrm{eff}$ respectively, with $\Delta \sigma$ and $\Delta R_\mathrm{eff}$ the observational uncertainties. We assume Gaussian uncertainties for simplicity, as our quantitative statements are only $1\sigma$--$2\sigma$ away from the best fit. We conservatively ignore correlations among $\Delta \sigma$ and $\Delta R_\mathrm{eff}$. Should correlations or non-Gaussian uncertainties be provided by observations, they would be straightforward to include. We consider halo masses between $10^7\, M_\odot$ and $10^{12}\, M_\odot$. Lower-mass halos would be too dim to be observed, as we discuss in \cref{sec:results_galaxy} (see also Ref.~\cite{Jethwa:2016gra}); and higher-mass halos would be heavier than the Milky Way.

The expression can be further simplified using conditional probabilities,
\begin{equation}
\label{eq:conditional_chain}
\begin{split}
 \mathcal{P}(\sigma_\mathrm{los}^*, R_\mathrm{eff}, c, M_*, M) & =  \mathcal{P}(\sigma_\mathrm{los}^*, R_\mathrm{eff}, c, M_* | M) \, \mathcal{P}(M) \\
 & = \mathcal{P}(\sigma_\mathrm{los}^*, R_\mathrm{eff} | c, M_*, M) \, \mathcal{P}(c, M_* | M) \, \mathcal{P}(M) \\
 & = \mathcal{P}(\sigma_\mathrm{los}^* | R_\mathrm{eff}, c, M_*, M)\, \mathcal{P}(R_\mathrm{eff} | c, M_*, M) \, \mathcal{P}(c, M_* | M) \, \mathcal{P}(M) \, ,
\end{split}
\end{equation}
where
\begin{itemize}
    \item $\mathcal{P}(M)$ is the product of the subhalo mass function at infall redshift (see \cref{sec:intro} in the main text) and the halo occupation fraction. We compute the former with \verb+Galacticus+ following Refs.~\cite{Sheth:2001dp, Benson:2019jio, Cole:2000ex, Parkinson:2007yh} as described in the main text (\cref{sec:consequences_DM_formalism}), and we parametrize the latter as described in the main text (\cref{eq:hof}, \cref{sec:satellites_formalism})
    \begin{equation}
    \operatorname{hof}(M) = \frac{1 + \operatorname{erf}\left(\alpha^\mathrm{hof} \log_{10}\left[M/M_0^\mathrm{hof}\right]\right)}{2} \, ,
    \end{equation}
    with erf the error function, and $\alpha^\mathrm{hof} > 1$~\cite{2019MNRAS.488.4585G, DES:2019ltu} and $M_0^\mathrm{hof}$ free parameters. $M$ is the infall mass~\cite{Barber:2013oua}.
    
    \item Following the observational results in Ref.~\cite{Kim:2021zzw}, we assume that $R_\mathrm{eff}$ only depends on stellar mass, i.e., ${\mathcal{P}(R_\mathrm{eff} | c, M_*, M) = \mathcal{P}(R_\mathrm{eff} | M_*)}$. As described in the main text (\cref{eq:R_eff}, \cref{sec:satellites_formalism}), we assume a lognormal distribution with median ${\overline{R_\mathrm{eff}} = 7.76\cdot 10^{-3} \, M_*^{0.268}}$ and scatter of 0.234\,dex~\cite{Kim:2021zzw}. 
    
    As $R_\mathrm{eff}$ only links to halo mass via the stellar mass-halo mass relation, that we set free (see below), observational uncertainties on $\mathcal{P}(R_\mathrm{eff} | M_*)$ do not have a strong impact on our error budget. We also set free the halo occupation fraction, which makes the connection between DM halos and $M_*$ even more flexible.

    \item We assume that $c$ and $M_*$ are independent and only related to $M$, $\mathcal{P}(c, M_* | M) = \mathcal{P}(c | M) \, \mathcal{P}(M_* | M)$. Then,
    
    \begin{itemize}
        \item $\mathcal{P}(c | M)$ is, as described in the main text (\cref{sec:consequences_DM_formalism}), a lognormal distribution with median computed with \texttt{Galacticus} following Ref.~\cite{Diemer:2018vmz} and scatter of 0.16\,dex~\cite{Diemer:2014gba}.
        \item $\mathcal{P}(M_* | M)$ is the product of the probability to observe the galaxy given its stellar mass, parametrized by the completeness correction; and the stellar mass-halo mass relation. 
        
        For the completeness correction, we follow \cref{eq:completeness1,eq:completeness2} in the main text (\cref{sec:satellites_formalism}). As explained in the main text, the completeness correction takes into account the dependence both on $M_*$ and $R_\mathrm{eff}$~\cite{Walsh:2008qn}.
        
        In principle, the completeness correction could be affected by low-surface-brightness uncertainties in the $M_*$--$R_\mathrm{eff}$ relation due to low selection efficiency. These have a minor impact on the \emph{shape} of the distribution of galaxies, i.e., $\mu(\sigma_\mathrm{los}^{*\,\mathrm{obs}}, R_\mathrm{eff}^\mathrm{obs})$; because they affect galaxies with a very low observation probability that are not included in our analysis (see, e.g., Fig.~4 in Ref~\cite{Nadler:2018iux}): we conservatively only include galaxies with an observation probability larger than 90\% (although the transition is rather sharp, see Ref.~\cite{Walsh:2008qn}). In turn, the \emph{total number} of galaxies, i.e., $\mathcal{P}(N_\mathrm{obs})$; could be affected more significantly. However, we have checked that low-surface-brightness uncertainties do not change the number of expected galaxies by more than $10\%$ (see also, e.g., Fig.~12 in Ref.~\cite{2022MNRAS.516.3944M}), which is a subdominant contribution to our error budget (see below).
        
        We include anisotropy-induced scatter as described in the main text (\cref{sec:analysis}): we multiply $\mathcal{C}_\Omega$ by a free parameter $\sigma_{\mathcal{C}_\Omega}$, and we add to the likelihood a Gaussian prior on $\sigma_{\mathcal{C}_\Omega}$ centered at 1 and with 19\% width~\cite{Tollerud:2008ze} (this is a conservative estimate, derived when SDSS was not yet complete). We also include uncertainty in the spatial satellite distribution $n(r)$ due to tidal disruption, by parametrizing $n(r)$ as described in the main text (\cref{eq:n_radial}, \cref{sec:analysis})
        \begin{equation}
            n(r) = n_\mathrm{GK17}(r) + y_\mathcal{C} [n_\mathrm{NFW}(r) - n_\mathrm{GK17}(r)] \, ,
        \end{equation}
        with $y_\mathcal{C} \in [0, 1]$ a free parameter that interpolates between an NFW distribution, $n_\mathrm{NFW}(r)$, and the tidally disrupted distribution from Ref.~\cite{Garrison-Kimmel:2017zes}, $n_\mathrm{GK17}(r)$. The latter distribution strongly suppresses the subhalo abundance within the inner $\sim 100 \, \mathrm{kpc}$. Enhancements in the number of subhalos within the inner $\sim 50 \, \mathrm{kpc}$ due to the recent passage of the Large Magellanic Cloud~\cite{Deason:2015hla, Barry:2023ksd, 2021MNRAS.504.5270D} are smaller than the tidal disruption uncertainty we parametrize by $y_\mathcal{C}$.
        
        For the stellar mass-halo mass relation, we follow \cref{eq:Mstar,eq:Mstar_scatter} in the main text (\cref{sec:satellites_formalism}). I.e., we assume it to be a lognormal distribution whose median follows a power law with power $\beta^{M_*}$ and normalization $M_*(M=1.54\times 10^{12} M_\odot) = 0.0455\,M$~\cite{Moster:2012fv}, and whose scatter can increase at low mass,~\cite{Garrison-Kimmel:2016szj}
        \begin{equation}
            \sigma(M) = \sigma^{M_*} + \gamma^{M_*} \log_{10} \frac{M}{10^{11} M_\odot} \,\mathrm{dex}\, .
        \end{equation}
        $\beta^{M_*}$, $\sigma^{M_*} \in [0,2]$~\cite{Garrison-Kimmel:2016szj, DES:2019ltu}, and $\gamma^{M_*} < 0$ are free parameters. $M$ is the infall mass~\cite{Moster:2012fv}.
    \end{itemize}
    \item $\mathcal{P}(\sigma_\mathrm{los}^* | R_\mathrm{eff}, c, M_*, M) = \delta\left( \sigma_\mathrm{los}^* - \sqrt{\frac{G}{4} \frac{M(<R_\mathrm{eff}/0.75)}{R_\mathrm{eff}}} \right)$ (\cref{eq:sigma_los} and \cref{sec:satellites_formalism} in the main text), which depends on $M$ and $c$ through the enclosed mass $M(<r)$. For simplicity we ignore scatter, as other mass estimators produce very similar results~\cite{Kim:2017iwr} (see also \cref{fig:sigma_estimator}). As described in the main text (\cref{sec:consequences_DM_formalism,sec:satellites_formalism}), we include baryonic feedback by assuming an NFW profile for $M < M_\mathrm{thres}^\mathrm{core}$ and a cored profile for $M > M_\mathrm{thres}^\mathrm{core}$ following Ref.~\cite{Read:2015sta} (where the core size depends on the time over which a galaxy has formed stars), with $M_\mathrm{thres}^\mathrm{core}$ a free parameter.
\end{itemize}
Since $\mu$ is a differential probability, we numerically normalize it so that $\int \mu(\sigma_\mathrm{los}^{*\,\mathrm{obs}}, R_\mathrm{eff}^\mathrm{obs}) \, \mathrm{d}\sigma_\mathrm{los}^{*\,\mathrm{obs}}\mathrm{d}R_\mathrm{eff}^\mathrm{obs} = 1$.

Finally, for the probability to observe $N_\mathrm{obs}$ satellite galaxies $\mathcal{P}(N_\mathrm{obs})$ we follow Ref.~\cite{Boylan-Kolchin:2009ztl} and include the additional scatter due to differences in the host halo's accretion histories~\cite{Fielder:2018szt} by modeling $\mathcal{P}(N_\mathrm{obs})$ with a negative binomial distribution with 18\% intrinsic scatter and mean
\begin{equation}
    N_\mathrm{expected} = \int \mathcal{P}(\sigma_\mathrm{los}^*, R_\mathrm{eff}, c, M_*, M) \, \mathrm{d}c\,\mathrm{d}M_* \,\mathrm{d}M \, \mathrm{d}\sigma_\mathrm{los}^* \, \mathrm{d}R_\mathrm{eff} \, ,
\end{equation}
i.e., $\mathcal{P}(\sigma_\mathrm{los}^*, R_\mathrm{eff}, c, M_*, M)$ is normalized to the expected number of observed galaxies. This integral is dominated by low-mass halos (see \cref{fig:hmf}), so the effect of not including the Magellanic clouds in our analysis is minor. However, the Large Magellanic Cloud accretes small-mass satellites with it, changing the expected number of satellites by up to tens of percent~\cite{Deason:2015hla, Barry:2023ksd, 2021MNRAS.504.5270D,DES:2019ltu,2021ApJ...920L..11N, 2022MNRAS.516.3944M} (although we use satellites discovered by SDSS, whose footprint does not contain the Magellanic Clouds). On top of that, changing the Milky Way mass within uncertainties~\cite{Cautun:2019eaf} would change $N_\mathrm{expected}$ by about 20\%~\cite{Kim:2021zzw}. Both effects are smaller than the total scatter of $\mathcal{P}(N_\mathrm{obs})$. In addition, our main constraining power on DM properties does not come from the total number of satellites, as that is degenerate with galaxy-halo connection parameters (see \cref{sec:results_galaxy} in the main text).

Overall, our likelihood $\mathcal{L}$ depends on 2 parameters describing DM physics, \{$n_\mathrm{cut}$, $k_\mathrm{cut}$\}; and 8 parameters modelling galaxy properties and baryonic effects, \{$M_0^\mathrm{hof}$, $\alpha^\mathrm{hof}$, $\beta^{M_*}$, $\sigma^{M_*}$, $\gamma^{M_*}$, $M_\mathrm{thres}^\mathrm{core}$, $\sigma_{\mathcal{C}_\Omega}$, $y_\mathcal{C}$\}. To obtain allowed regions in a subset of parameters, we minimize $-2 \ln \mathcal{L}$ over other parameters. We draw contours corresponding to $\Delta 2 \ln \mathcal{L} = 1.0$ and $4.0$ for one-parameter scans; and to $\Delta 2 \ln \mathcal{L} = 2.30$ and $6.18$ for two-parameter scans. Under the assumption that Wilks' theorem~\cite{Wilks:1938dza} holds, these can be interpreted as 68\% (1$\sigma$) and 95\% (2$\sigma$) confidence level regions. We note that this theorem requires the true values to be well-contained within the parameter space range and a large sample size; and we use broad scan ranges (see \cref{sec:app_full_results} for a discussion and visualization of the results), plus the amount of data (21 values of $\sigma_\mathrm{los}^*$, 21 values of $R_\mathrm{eff}$, and the total number of galaxies) is relatively large. We carry out the integrals with the \verb+vegas+ package~\cite{Lepage:2020tgj} and the minimizations with the \verb+Py-BOBYQA+ package~\cite{bobyqa, 10.1145/3338517}.

\clearpage
\setcounter{figure}{0}
\section{Full parameter space scan}
\label{sec:app_full_results}

In this Appendix, we provide the full results of our statistical analysis.

\Cref{tab:results} shows our free parameters, the ranges over which we scan in our analysis (in a Bayesian framework, these would correspond to the prior ranges), and the $1\sigma$ preferred ranges after minimizing $-2 \ln \mathcal{L}$ over all other parameters. We find that we can constrain many galaxy-halo connection parameters, such as the slope of the stellar mass-halo mass relation, the turning mass of the halo occupation fraction, the halo mass above which baryonic feedback makes DM density profile cored, or the amount of subhalo tidal disruption. Other parameters, such as the slope of the halo occupation fraction or the low-mass growth of stellar mass-halo mass scatter, are more challenging to constrain.

\begin{table}[hb]
    \centering
    \begin{tabular}{c|c|c|c|c|c}
    & $k_\mathrm{cut}/\mathrm{Mpc^{-1}}$ & $n_\mathrm{cut}$ & $M_0^\mathrm{hof} / M_\odot$ & $\alpha^\mathrm{hof}$ & $\beta^{M_*}$ \\\hline\rule{0pt}{\normalbaselineskip}

    \multirow{2}{*}{Meaning} & Scale above which & Slope of the & Halo mass above which & Steepness of the halo & Slope of the stellar mass-\\\rule{0pt}{0pt}
    & $P(k)$ is enhanced & enhanced $P(k)$ & halos host galaxies & occupation fraction& halo mass relation\\\rule{0pt}{\normalbaselineskip}
    Definition & \cref{eq:powerSpec}& \cref{eq:powerSpec}& \cref{eq:hof}& \cref{eq:hof}& \cref{eq:Mstar} \\\rule{0pt}{\normalbaselineskip}
    Value in \cref{fig:data} & 8 & 2.6 & $10^{8.35}$~\cite{Barber:2013oua, Dooley:2016xkj}&  $1.31$~\cite{Barber:2013oua, Dooley:2016xkj}& 0.963~\cite{Moster:2012fv}\\\rule{0pt}{\normalbaselineskip}
    Scan range & (4, 45) & (1, 5)& ($10^7$, $10^{11}$)& (1, 10)~\cite{2019MNRAS.488.4585G, DES:2019ltu} & (0, 3)\\\rule{0pt}{\normalbaselineskip}
    $1\sigma$ range &(4, 45)$^*$ &(1, 5)* & ($10^7$, $10^{7.9}$)& (1, 10)& (1.0, 1.7)\\ 
    \end{tabular}

    \vspace*{0.5cm}

    \centerline{
    \begin{tabular}{c|c|c|c|c|c}
    & $\sigma^{M_*}$/dex & $\gamma^{M_*}$/dex & $M_\mathrm{thres}^\mathrm{core} / M_\odot$ & $y_\mathcal{C}$ & $\sigma_{\mathcal{C}_\Omega}$ \\\hline\rule{0pt}{\normalbaselineskip}
    \multirow{2}{*}{Meaning} & Stellar mass-halo mass& Mass dependence of& Halo mass below which & Amount of tidal & Anisotropy in the \\\rule{0pt}{0pt}
    &scatter at $10^{11}\,M_\odot$ &stellar mass-halo mass scatter& DM profiles are cored& halo disruption& satellite distribution\\\rule{0pt}{\normalbaselineskip}    
    Definition & \cref{eq:Mstar_scatter} & \cref{eq:Mstar_scatter}& Below \cref{eq:sigma_los}& \cref{eq:n_radial} & Below \cref{eq:n_radial}\\\rule{0pt}{\normalbaselineskip}
    Value in \cref{fig:data} & 0.15~\cite{Moster:2012fv}& 0~\cite{Moster:2012fv} & $10^9\,M_\odot$~\cite{Kim:2021zzw}& 1 & 1\\\rule{0pt}{\normalbaselineskip}
    Scan range & (0, 2)~\cite{Garrison-Kimmel:2016szj, DES:2019ltu}& (-2, 0)& ($10^7$, $10^{11}$) & (0, 1)& (0, 2)\\\rule{0pt}{\normalbaselineskip}
    $1\sigma$ range & (0, 1.5)& (-0.6, -0.2)& ($10^7$, $10^{7.2}$)& (0.4, 1) & (0.87, 1.14)\\ 
    \end{tabular}
    }
    \caption{Parameters in our analysis. The parameters whose $1\sigma$ range has a $^*$ are strongly correlated with other parameters (see \cref{fig:triangle}); meaningful constraints can be derived when other parameters are fixed. For the theoretical population in \cref{fig:data}, we use illustrative galaxy-halo connection parameters (see references). Some scan ranges are physics-motivated (see references).}
    \label{tab:results}
\end{table}

\begin{figure}[ht]
    \centering
    \includegraphics[width=\textwidth]{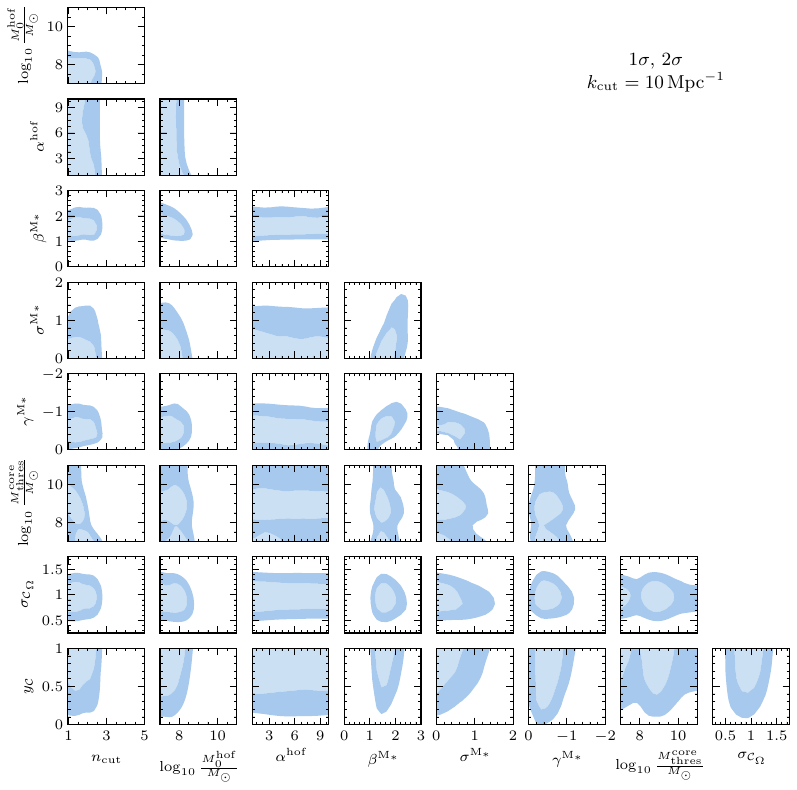}
   
    \caption{Allowed regions within $1\sigma$ and $2\sigma$ for all parameters in our analysis. For each pair of parameters shown, all other parameters are minimized over in the likelihood. To visualize correlations with our determination of the power spectrum, we fix ${k_\mathbf{cut} = 10 \, \mathrm{Mpc^{-1}}}$. \emph{Parameters controlling galaxy-halo connection and baryonic feedback are mostly independent of the DM power spectrum}.}
    \label{fig:triangle}
\end{figure}

\Cref{fig:triangle} shows the correlations among our free parameters. To better visualize correlations between galaxy-halo connection parameters and enhanced power spectra (i.e., $n_\mathrm{cut} > 1$), we fix $k_\mathrm{cut}=10\,\mathrm{Mpc^{-1}}$. Otherwise, for large enough $k_\mathrm{cut}$, any value of $n_\mathrm{cut}$ would be degenerate in our halo mass range with a scale-invariant power spectrum (see \cref{fig:physics,fig:LDM_constraints}). We show the preferred $1\sigma$ and $2\sigma$ regions after minimizing $-2 \ln \mathcal{L}$ over all other parameters. We find that, even though there are correlations among galaxy-halo connection parameters --- particularly for the stellar mass-halo mass relation ---, enhanced power, i.e., a modified $n_\mathrm{cut}$, is mostly independent of galaxy-halo connection as described in the main text (\cref{sec:results_galaxy}). As discussed in \cref{sec:results_DM}, the main correlation of enhanced power is with the halo mass above which baryonic feedback makes DM density profiles cored, i.e., $M_\mathrm{thres}^\mathrm{core}$. This makes our main conclusions robust even though for some parameters the $1\sigma$ range is bounded by the scan range. In more detail, our analysis does not fully bound the following parameters,

\begin{itemize}
    \item $M_0^\mathrm{hof}$ and $M_\mathrm{core}^\mathrm{thres}$: both the 1$\sigma$ range and the scan range stop at $10^7\,M_\odot$. As discussed in \cref{sec:app_statistics}, this corresponds to the smallest halo mass we consider in our analysis, so smaller values of $M_0^\mathrm{hof}$ and/or $M_\mathrm{core}^\mathrm{thres}$ would produce the same results.

    \item $y_\mathcal{C}$: both the 1$\sigma$ range and the scan range stop at $1$. As discussed in \cref{sec:app_statistics}, $y_\mathcal{C}=1$ corresponds to no tidal halo disruption, so this is a physical boundary.

    \item $\alpha^\mathrm{hof}$: the 1$\sigma$ range covers the whole scan range. From \cref{eq:hof}, large $\alpha^\mathrm{hof}$ makes the transition in the halo occupation fraction more steep. At $\alpha^\mathrm{hof}=10$ the transition is already a step function for all purposes, so higher $\alpha^\mathrm{hof}$ would produce the same result. In turn, $\alpha^\mathrm{hof}<1$ would make the halo occupation fraction so flat that very heavy ($M \gtrsim 10^{11} M_\odot$) dark halos would be predicted, against simulations (see, e.g., Ref.~\cite{2019MNRAS.488.4585G}) and observations.

    \item $\sigma^{M_*}$: both the $1\sigma$ range and the scan range stop at 0. This corresponds to the physical boundary of scatter being positive.
\end{itemize}

\clearpage
\setcounter{figure}{0}
\section{Data used in the analysis}
\label{sec:app_data}

For reference, in this Appendix we compile the data used in our analysis, as obtained from Ref.~\cite{McConnachie:2012vd}. As in Ref.~\cite{Kim:2021zzw}, for Bootes II we adopt $\sigma_\mathrm{los}^* = 4.4 \pm 1 \, \mathrm{km/s}$, instead of $\sigma_\mathrm{los}^* = 10.5 \pm 7.4 \, \mathrm{km/s}$ as listed in Ref.~\cite{McConnachie:2012vd}. As discussed in Ref.~\cite{Kim:2021zzw}, our value is based on a determination that employs more stars.

\begin{figure}[b]
    \centering
    \includegraphics[width=0.45\textwidth]{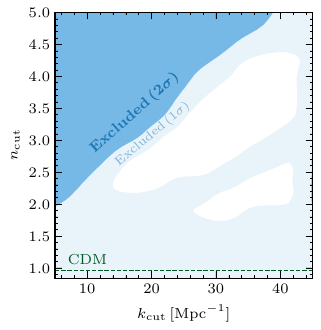}
    \caption{Excluded values of LDM parameter space in our analysis, if the original determination of $\sigma_\mathrm{los}^*$ for Segue II is used.}
    \label{fig:LDM_constraints_Segue}
\end{figure}

In addition, for Segue II the original determination quoted in Ref.~\cite{McConnachie:2012vd}, $\sigma_\mathrm{los}^* = 3.4^{+2.5}_{-1.2}\,\mathrm{km/s}$~\cite{Belokurov:2009hw}, was later reanalyzed, leading only to an upper limit $\sigma_\mathrm{los}^* < 1.4\,\mathrm{km/s}$ (68\% CL)~\cite{Kirby:2013isa}. \Cref{fig:LDM_constraints_Segue} shows the excluded values of LDM parameters if we carry out our analysis with the original determination. Comparing with \cref{fig:LDM_constraints}, where Segue II is included as an upper limit, we see that the results are similar in both scenarios. Since the measurements leading to the upper limit contain a larger spectroscopic sample, and the resulting constraint on LDM models is more conservative; we have chosen $\sigma_\mathrm{los}^* < 1.4\,\mathrm{km/s}$ for our main analysis. We note that, even though the low velocity dispersion of Segue II may be the outcome of significant tidal stripping, it is only a $\sim 2\sigma$ outlier from the overall trend in \cref{fig:data}. Such outliers are to be expected given that we consider $\sim 20$ galaxies. Furthermore, the similar results obtained in our analysis with the original determination and with the upper limit emphasize that our population study is sensitive to \emph{average} galaxy properties. Hence, it is reasonably robust against outliers.

We implement galaxies with asymmetric error bars in \cref{eq:likelihood} by using asymmetric split-Gaussian distributions.
\begin{table}[hbtp]
    \setlength{\tabcolsep}{5pt}
\parbox[t]{.49\linewidth}{
\begin{center}
    \begin{tabular}{ccc}
    \toprule
    Galaxy & $\sigma_\mathrm{los}^*$ [km/s] & $R_\mathrm{eff}$ [kpc] \\
    \midrule
    Fornax & $11.7 \pm 0.9$ & $0.710 \pm 0.077$ \\
    Leo I & $9.2 \pm 1.4$ & $0.251 \pm 0.027$ \\
    Sculptor & $9.2 \pm 1.4$ & $0.283 \pm 0.045$ \\
    Leo II & $6.6 \pm 0.7$ & $0.176 \pm 0.042$ \\
    Sextans I & $7.9 \pm 1.3$ & $0.695 \pm 0.044$ \\
    Carina & $6.6 \pm 1.2$ & $0.250 \pm 0.039$ \\
    Draco & $9.1 \pm 1.2$ & $0.221 \pm 0.019$ \\
    Ursa Minor & $9.5 \pm 1.2$ & $0.181 \pm 0.027$ \\
    Canes Venatici I & $7.6 \pm 0.4$ & $0.564 \pm 0.036$ \\
    Hercules & $3.7 \pm 0.9$ & $0.330^{+0.075}_{-0.052}$\\
    
    \end{tabular}
    \end{center}
    } \parbox[t]{.45\textwidth}{
    \begin{center}
\begin{tabular}{ccc}
    \toprule
    Galaxy & $\sigma_\mathrm{los}^*$ [km/s] & $R_\mathrm{eff}$ [kpc] \\
    \midrule
    Bootes I & $2.4^{+0.9}_{-0.5}$ & $0.242 \pm 0.021$ \\
    Leo IV & $3.3 \pm 1.7$ & $0.206 \pm 0.037$ \\
    Ursa Major I & $7.6 \pm 1.0$ & $0.319 \pm 0.050$ \\
    Leo V & $3.7_{-1.4}^{+2.3}$ & $0.135 \pm 0.032$ \\
    Canes Venatici II & $4.6 \pm 1.0$ & $0.074 \pm 0.014$ \\
    Ursa Major II & $6.7 \pm 1.4$ & $0.149 \pm 0.021$ \\
    Coma Berenices & $4.6 \pm 0.8$ & $0.077 \pm 0.010$ \\
    Bootes II & $4.4 \pm 1.0$ & $0.051 \pm 0.017$ \\
    Willman 1 & $4.3_{-1.3}^{+2.3}$ & $0.025 \pm 0.006$ \\
    Segue II & $<1.4$ & $0.035 \pm 0.003$ \\
    Segue I & $3.9 \pm 0.8$ & $0.029_{-0.005}^{+0.008}$\\
    
    \end{tabular}
    \end{center}
    }
    \caption{List of data used in the analysis}
\end{table}

\end{document}